\documentclass[showpacs,prb]{revtex4}
\usepackage{amssymb}
\usepackage{amsmath}
\usepackage{graphicx}

\begin{document}

\preprint{}
\title{Spin dynamics in a doped-Mott-insulator superconductor}
\author{W.Q. Chen and Z.Y. Weng}
\affiliation{Center for Advanced Study, Tsinghua University, Beijing 100084, China}

\begin{abstract}
We present a systematic study of spin dynamics in a superconducting ground
state, which itself is a doped-Mott-insulator and can correctly reduce to an
antiferromagnetic (AF) state at half-filling with an AF long-range order
(AFLRO). Such a doped Mott insulator is described by a mean-field theory
based on the phase string formulation of the $t-J$ model. We show that the
well-known spin wave excitation in the AFLRO state at half-filling evolves
into a resonancelike peak at a finite energy in the superconducting state,
which is located around the AF wave vectors. The width of such a
resonancelike peak in momentum space decides a spin correlation length scale
which is inversely proportional to the square root of doping concentration,
while the energy of the resonancelike peak scales linearly with the doping
concentration at low doping. These properties are consistent with
experimental observations in the high-$T_{c}$ cuprates. An important
prediction of the theory is that, while the total spin sum rule is satisfied
at different doping concentrations, the weight of the resonancelike peak
does not vanish, but is continuously saturated to the weight of the AFLRO at
zero-doping limit. Besides the low-energy resonancelike peak, we also show
that the high-energy excitations still track the spin wave dispersion in
momentum space, contributing to a significant portion of the total spin sum
rule. The fluctuational effect beyond the mean-field theory is also
examined, which is related to the broadening of the resonancelike peak in
energy space. In particular, we discuss the incommensurability of the spin
dynamics by pointing out that its visibility is strongly tied to the
low-energy fluctuations below the resonancelike peak. We finally investigate
the interlayer coupling effect on the spin dynamics as a function of doping,
by considering a bilayer system.
\end{abstract}

\maketitle

\section{\protect\bigskip Introduction}

\label{sec:intro}

The measurement of spin dynamics in the cuprate superconductors is uniquely
important. This is because the spin degrees of freedom constitute the
predominant part of the low-lying electronic degrees of freedom, \emph{i.e.,
}$1-\delta $ per site, as compared to the charge degrees of freedom at small
hole concentration, $\delta $ per site. Such a large imbalance between the
spin and charge numbers are usually regarded as a key indicator that the
underlying system is a doped Mott insulator \cite{and}. On general grounds,
the corresponding spin dynamics is expected to be distinctly different from
a conventional BCS superconductor. The latter is based on the Fermi-liquid
description in which the elementary excitations are quasiparticles that
carry both charge and spin. An extreme case is at half-filling, where the
whole charge degrees of freedom get frozen at low energy and only the spin
degrees of freedom remain intact in the cuprates, whose dynamics is well
characterized by the\ Heisenberg model \cite{chak}.

Experimentally, anomalous properties of spin dynamics have been observed
throughout the cuprate family. The parent compound at half-filling is a Mott
insulator in which spins form AFLRO below a N\'{e}el temperature $T_{N}$.
The elementary excitation is a gapless bosonic Goldstone mode, \emph{i.e.,}
the spin wave in the ordered phase. AFLRO and the spin-wave excitation
disappear beyond some critical concentration of holes introduced into the
system. Except for some residual signature of spin waves at high energies,
the low-lying spin-wave-type excitation is completely absent once the system
becomes a superconductor. It is replaced by a resonancelike peak at a
doping-dependent energy around the AF wave vector $\mathbf{Q}_{\text{\textrm{%
AF}}}=(\pi ,\pi )$, as observed first in the optimally doped YBCO compound
\cite{FirstDiscover,PRL75_316}, where the dynamic spin susceptibility
function measured by inelastic neutron scattering shows a sharp peak at $%
\omega _{\text{\textrm{res}}}=41$ \textrm{meV}$,$ whose width is comparable
to the resolution limit of the instruments. Similar resonancelike peak has
also been observed in the underdoped YBCO compounds \cite{UnderdopedYBCO}
(where the resonancelike peak persists into the pseudogap phase above the
superconducting transition), Tl-based \cite{TBCO} and Bi-based \cite{BSCCO}
compounds. In the LSCO compound, although no such a sharp peak has been
found, the low-lying spin excitation is nonetheless non spin-wave-like,
which may be still regarded as a very broad peak in energy space \cite%
{hayden}. With much sharper linewidth in momentum space, doping-dependent
incommensurate splittings around $\mathbf{Q}_{\text{\textrm{AF}}}$ have been
clearly identified in LSCO \cite{LSCO,yamada}. Similar incommensurability,
even though not as prominent as in LSCO, has been also established in
underdoped \textrm{YBCO} recently \cite{YBCOInc1, YBCOInc2, YBCOInc3}.

Theoretically, a great challenge is how to naturally connect the spin
dynamic at half-filling with that in the superconducting phase in which the
doping concentration can be as low as a few percent per\ \textrm{Cu }site.
That is, although the low-energy, long-wavelength behavior may change
qualitatively in the superconducting phase, the number of spins in the
background is still quite close to half-filling, which far exceeds the
number of doped holes. Physically it is very hard to imagine that the \emph{%
short-range, high-energy} spin correlations would be changed completely by a
few percent to ten percent doping. However, in a BCS superconductor, the
upper spin energy scale is usually set by the Fermi energy $\epsilon _{f}$
\cite{si}, such that in the local spin susceptibility one has to integrate
over the frequency up to $\epsilon _{f}$ in order to recover the correct sum
rule of $1-\delta $ spin per site. Normally $\epsilon _{f}$ is much larger
than $J$. Thus, why there should be a gigantic increase in the upper spin
energy in the doped case, compared to the half-filling, poses a serious
challenge to any approach based on the d-wave BCS-type theory in which the
spin dynamic is solely contributed by quasiparticle excitations.
Experimentally the upper energy scale exhibited in the dynamic spin
susceptibility is set by $\sim 2J$ ($J$ is the superexchange coupling) at
half-filling, in consistency with the prediction by the Heisenberg model,
and is slightly reduced in the optimal-doped superconducting phase \cite%
{hayden}. No trace of any other new high-energy scale has been ever reported
in the doped regime in spin channels.

As for the low-energy feature, like the resonancelike peak structure
observed in the experiments, theoretical proposals are ranged from the RPA
fluctuations in the particle-hole channel within the framework of BCS \cite%
{si} or generalized BCS theories \cite{lan,brinkman} to some novel mechanism
of the so-called $\pi $ mode in the particle-particle channel in the SO(5)
theory \cite{so5}, which is coupled to the particle-hole channel in the
superconducting phase. An important question, not being properly addressed
yet, is what is the connection, if any, of such a resonancelike spin mode
with the spin wave in the zero-doping limit. Namely, how a few percent of
doped holes can continuously reshape a spin-wave excitation into a
non-propagating local mode, with an AFLRO turning into short-range spin
correlations. This question and the previous high-energy one constitute two
of most fundamental issues in an approach based on doped Mott insulators.
\emph{\ }

In this paper, we put forward a systematic description of the evolution of
spin dynamics as a function of doping in a doped-Mott-insulator
superconductor. It is described by a bosonic resonating-valence-bond (RVB)
mean-field theory \cite{phasestring_meanfield} based on the phase-string
formulation \cite{phasestring_general} of the $t-J$ model. At half-filling,
the mean-field theory reduces to the Schwinger-boson mean-field state \cite%
{schwinger_boson}, which well characterizes AFLRO and spin-wave excitations
in the ground state. At finite doping, the mean-field theory depicts how the
spin dynamics is influenced by the doping effect in going into the
superconducting state. In particular, we show how a resonancelike peak
centered around $\mathbf{Q}_{\text{\textrm{AF}}}$ emerges out of spin waves
from the AFLRO phase. A unique prediction for experiment is that the weight
of the resonancelike peak continuously evolves into that of the AFLRO in the
zero-doping limit. On the other hand, the total weight of the dynamic
susceptibility function, which extends slightly over $\sim 2J$ in energy,
still satisfies the sum rule that the total spin number is $1-\delta $ per
site.

In this unified mean-field description, doping-dependent resonancelike
energy and spin correlation length are quantitatively determined. Besides
the low-energy resonancelike peak structure near $\mathbf{Q}_{\text{\textrm{%
AF}}}$, there still exists a high-energy spectrum whose envelope roughly
tracks the spin wave dispersion as a residual effect in the superconducting
phase. We also consider some leading fluctuational effect beyond the
mean-field theory on the lineshape of the spectral function, and discuss the
incommensurability and its visibility in this framework. We finally
introduce the interlayer superexchange coupling and investigate how the spin
dynamics changes in the even and odd channels for a double-layer system.
Comparisons with the experimental measurements, mostly by inelastic neutron
scattering, are made.

The remainder of the paper is organized as follows. In Sec. II, a systematic
study of spin dynamics in the bosonic RVB mean-field state for the
single-layer system is presented. In Sec. III, fluctuational effects beyond
the mean-field theory, due to the charge density fluctuations, are
discussed. In Sec. IV, the interlayer coupling for a bilayer system is
considered. Finally, a summary is given in Sec. V.

\section{Spin Dynamics in Mean Field Description}

\label{sec:psmfa}

\subsection{Bosonic RVB state at half-filling}

Spin dynamics of the cuprates at half-filling is well described by the
two-dimensional (2D)\ AF Heisenberg model. Although a conventional spin-wave
theory is quite successful in understanding the low-lying excitation
spectrum of the Heisenberg Hamiltonian, to make the theory applicable or
modifiable to the cases without AFLRO, like at finite temperatures or in
doped regimes, we shall use the Schwinger-boson formulation as our starting
point at half-filling.

The mean-field theory \cite{schwinger_boson} based on the Schwinger-boson
formulation can characterize the AFLRO and spin-wave excitation fairly well
in the ground state. Its mean-field wavefunction under the Gutzwiller
projection will have the same form \cite{weng_04} as the variational bosonic
RVB wavefunctions proposed by Liang, Doucot, and Anderson \cite%
{LiangDoucotAnderson}. The latter can produce very accurate variational
energies as well as the AF\ magnetization for the Heisenberg model,
indicating that the state correctly captures \emph{both} long-range and
short-range spin correlations. Such an approach is thus called bosonic RVB
description, which is to be generalized to finite doping in the next
subsection. In the following, we briefly review some basic equations of the
bosonic RVB mean-field theory at half-filling.

In the Schwinger-boson formulation, the spin operators can be expressed by
the Schwinger-boson operator $b_{i\sigma }$ as follows

\begin{equation}
S_{i}^{+}=(-1)^{i}b_{i\uparrow }^{\dagger }b_{i\downarrow },  \label{s+}
\end{equation}%
(note that a staggered sign factor $(-1)^{i}$ is explicitly introduced here
in contrast to the original definition \cite{schwinger_boson}), and $%
S_{i}^{-}=(S_{i}^{+})^{\dagger },$ while $S_{i}^{z}=\sum_{\sigma }\sigma
b_{i\sigma }^{\dagger }b_{i\sigma }.$ The Schwinger bosons satisfy the
constraint $\sum_{\sigma }b_{i\sigma }^{\dagger }b_{i\sigma }=1.$ The
mean-field state is characterized by the bosonic RVB order parameter

\begin{equation}
\Delta _{0}^{s}=\sum_{\sigma }\langle b_{i\sigma }b_{j-\sigma }\rangle ,
\label{brvb0}
\end{equation}%
which leads to the following effective Hamiltonian, obtained from the
half-filling $t-J$ (Heisenberg) model:
\begin{equation}
H_{s}=-\frac{J\Delta _{0}^{s}}{2}\sum_{\langle ij\rangle \sigma }b_{i\sigma
}^{\dagger }b_{j-\sigma }^{\dagger }+H.c.+\text{ }\mathrm{const.}+\lambda
\left( \sum_{i\sigma }b_{i\sigma }^{\dagger }b_{i\sigma }-N\right) ,
\label{hs0}
\end{equation}%
where the last term involves a Lagrangian multiplier $\lambda $ to enforce
the global constraint of total spinon number, $\sum_{i\sigma }b_{i\sigma
}^{\dagger }b_{i\sigma }=N$.

The mean-field Heisenberg Hamiltonian (\ref{hs0}) can be straightforwardly
diagonalized by the Bogoliubov transformation
\begin{equation}
b_{i\sigma }=\sum_{\mathbf{k}}\omega _{\mathbf{k}\sigma }(i)(u_{\mathbf{k}%
}\gamma _{\mathbf{k}\sigma }-v_{\mathbf{k}}\gamma _{\mathbf{k}-\sigma
}^{\dagger }),  \label{bogo}
\end{equation}%
as
\begin{equation}
H_{s}=\sum_{\mathbf{k}\sigma }E_{\mathbf{k}}\gamma _{\mathbf{k}\sigma
}^{\dagger }\gamma _{\mathbf{k}\sigma }.
\end{equation}%
Here, $\omega _{\mathbf{k}\sigma }(i)=\frac{1}{\sqrt{N}}e^{i\sigma \mathbf{%
k\cdot r}_{i}},$ and the coherent factors, $u_{\mathbf{k}}$ and $v_{\mathbf{k%
}},$ are given by
\begin{equation}
u_{\mathbf{k}}=\frac{1}{\sqrt{2}}\sqrt{\frac{\lambda }{E_{\mathbf{k}}}+1}%
,\quad v_{k}=\frac{\mathrm{sgn}(\xi _{\mathbf{k}})}{\sqrt{2}}\sqrt{\frac{%
\lambda }{E_{\mathbf{k}}}-1},
\end{equation}%
where $\xi _{\mathbf{k}}=-J\Delta _{0}^{s}(\cos k_{x}a+\cos k_{y}a)$ and
\begin{equation}
E_{\mathbf{k}}=\sqrt{\lambda ^{2}-\xi _{\mathbf{k}}^{2}}.
\end{equation}%
Finally, in a self-consistent manner, the RVB order parameter $\Delta
_{0}^{s}$ and the Lagrangian multiplier $\lambda $ are determined by the
following self-consistent equations
\begin{align}
& \left| \Delta _{0}^{s}\right| ^{2}=\frac{1}{2N}\sum_{\mathbf{k}}\frac{\xi
_{\mathbf{k}}^{2}}{JE_{\mathbf{k}}}\coth \frac{\beta E_{\mathbf{k}}}{2}, \\
& 2=\frac{1}{N}\sum_{\mathbf{k}\neq 0}\frac{\lambda }{E_{\mathbf{k}}}\coth
\frac{\beta E_{\mathbf{k}}}{2}+n_{BC}^{b},
\end{align}%
in which $n_{BC}^{b}$ is the contribution from the Bose condensation of the
Schwinger bosons, leading to an AFLRO, which happens if $E_{\mathbf{k}}$
becomes gapless. Note that $\beta =1/T$ and the AFLRO disappears ($%
n_{BC}^{b}=0$) at a finite temperature $T$.

\subsection{Bosonic RVB description at finite doping}

Although AF correlations at half-filling are well captured by the mean-field
Hamiltonian $H_{s}$ in (\ref{hs0}), the doping effect on the spin background
is a highly nontrivial issue.

Based on the phase-string formulation \cite{phasestring_general}, which is
an exact reformulation by sorting out the most singular doping effect, \emph{%
i.e.,} the phase string effect in the $t-J$ model, a generalized mean-field
Hamiltonian describing the spin degrees of freedom can be obtained \cite%
{phasestring_meanfield} as follows

\begin{equation}
H_{s}=-\frac{J\Delta ^{s}}{2}\sum_{\langle ij\rangle \sigma }b_{i\sigma
}^{\dagger }b_{j-\sigma }^{\dagger }e^{i\sigma A_{ij}^{h}}+H.c.+\text{ }%
\mathrm{const.}+\lambda \left( \sum_{i\sigma }b_{i\sigma }^{\dagger
}b_{i\sigma }-(1-\delta )N\right) .  \label{hs}
\end{equation}

Compared to the half-filling case, $H_{s}$ in (\ref{hs}) differs from (\ref%
{hs0}) mainly by the emergence of a gauge field $A_{ij}^{h}$ defined on the
nearest-neighboring (NN) link $(ij),$ satisfying the following constraint%
\begin{equation}
\sum_{\langle ij\rangle \in c}A_{ij}^{h}=\pi \sum_{l\in \Omega
_{c}}n_{l}^{h},  \label{ah}
\end{equation}%
where $c$ is a, say, counter-clockwise-oriented close loop and $\Omega _{c}$
is the area enclosed by $c.$ On the right-hand-side (rhs) of (\ref{ah}), $%
n_{l}^{h}$ denotes the number operator of doped holes at site $l$.
Therefore, the doping effect explicitly enters in (\ref{hs}) through the
gauge field $A_{ij}^{h}$ as if each hole carries a fictitious $\pi $ fluxoid
as seen by spinons in $H_{s}$. In (\ref{hs}), the bosonic RVB\ order
parameter is given by
\begin{equation}
\Delta ^{s}=\sum_{\sigma }\left\langle e^{-i\sigma A_{ij}^{h}}b_{i\sigma
}b_{j-\sigma }\right\rangle _{NN}  \label{brvb}
\end{equation}%
for NN sites $i$ and $j$. At half filling, because there is no hole, it is
obvious that $A_{ij}^{h}=0$, and $\Delta ^{s}$ reduces back to $\Delta
_{0}^{s}$ defined in (\ref{brvb0}).

Note that the doping concentration $\delta $\ also enters $H_{s}$ through
the Lagrangian multiplier $\lambda $ which implements the global condition $%
\sum_{i\sigma }b_{i\sigma }^{\dagger }b_{i\sigma }=(1-\delta )N$. But at low
doping, the effect of missing spins represented by such a term will be far
less dramatic than the topological gauge field $A_{ij}^{h}.$ The latter
reflects the singular phase string effect \cite{phasestring_general} induced
by the hopping of doped holes on the AF spin background.

Corresponding to $H_{s}$ in (\ref{hs}), the spin operators in the phase
string formulation \cite{phasestring_general} read%
\begin{equation}
S_{i}^{+}=(-1)^{i}e^{i\Phi _{i}^{h}}b_{i\uparrow }^{\dagger }b_{i\downarrow
},  \label{s+1}
\end{equation}%
$S_{i}^{-}=(S_{i}^{+})^{\dagger },$ and $S_{i}^{z}=\sum_{\sigma }\sigma
b_{i\sigma }^{\dagger }b_{i\sigma },$ respectively$.$ Compared to the
Schwinger-boson formulation in (\ref{s+}), an extra phase $\Phi _{i}^{h}$
appears in (\ref{s+1}), which satisfies $\Phi _{i}^{h}-\Phi
_{j}^{h}=2A_{ij}^{h}$ ($ij\in $ spin sites) and ensures the spin rotational
symmetry of the effective Hamiltonian (\ref{hs}).

Equation (\ref{hs}) is by nature a gauge model. But in the superconducting
ground state, due to the Bose condensation of bosonic holons in the bosonic
RVB theory \cite{phasestring_meanfield}, the spin Hamiltonian $H_{s}$ will
become quite simplified as $A_{ij}^{h}$ can be approximately treated as
describing a uniform flux with a strength

\begin{equation}
\sum\nolimits_{\square }A_{ij}^{h}=\pi \delta .  \label{uflux}
\end{equation}

Then we can introduce the following Bogoliubov transformation to diagonalize
(\ref{hs}), just like (\ref{bogo}) in diagonalizing (\ref{hs0}),
\begin{equation}
b_{i\sigma }=\sum_{m}\omega _{m\sigma }(i)(u_{m}\gamma _{m\sigma
}-v_{m}\gamma _{m-\sigma }^{\dagger }).  \label{eq:2}
\end{equation}%
With a standard procedure, we obtain
\begin{equation}
H_{s}=\sum_{m\sigma }E_{m}\gamma _{m\sigma }^{\dagger }\gamma _{m\sigma },
\end{equation}%
with the spinon spectrum
\begin{equation}
E_{m}=\sqrt{\lambda ^{2}-\xi _{m}^{2}}.  \label{em}
\end{equation}%
In this scheme, $\xi _{m}$ and $\omega _{m\sigma }(i)=\omega _{m-\sigma
}^{\ast }(i)$ are eigenvalues and eigenfunctions of the following equation
\begin{equation}
\xi _{m}\omega _{m\sigma }(i)=-\frac{J\Delta ^{s}}{2}\sum_{j=NN(i)}e^{i%
\sigma A_{ij}^{h}}\omega _{m\sigma }(j).  \label{eq:3}
\end{equation}%
and the coherent factors, $u_{m}$ and $v_{m},$ are given by
\begin{equation}
u_{m}=\frac{1}{\sqrt{2}}\sqrt{\frac{\lambda }{E_{m}}+1},\quad v_{m}=\mathrm{%
sgn}(\xi _{m})\frac{1}{\sqrt{2}}\sqrt{\frac{\lambda }{E_{m}}-1}.
\end{equation}%
Finally, $\lambda $ and $\Delta ^{s}$ can be determined by the
self-consistent equations
\begin{align}
& \left\vert \Delta ^{s}\right\vert ^{2}=\frac{1}{2NJ}\sum_{m}\frac{\xi
_{m}^{2}}{E_{m}}\coth \frac{\beta E_{m}}{2},  \label{ds} \\
& 2-\delta =\frac{1}{N}\sum_{m\neq 0}\frac{\lambda }{E_{m}}\coth \frac{\beta
E_{m}}{2}+n_{BC}^{b}.  \label{lambda}
\end{align}%
Here $n_{BC}^{b}$ is the contribution of the condensation of spinons, if an
AFLRO\ exists like in the half-filling case.

The above mean-field formulation is essentially the same as the one outlined
in Ref. \cite{phasestring_meanfield}. For simplicity and clarity, here we
have not explicitly included an approximate doping-correction factor in (\ref{ds}) ($%
\approx 1-2\delta $) as we will be mainly concerned with the
evolution of spin dynamics at low doping. Such additional
corrections from doped holes can be always incorporated by simply
replacing the superexchange coupling $J$ with a doping-dependent
$J_{eff}$ which is quickly reduced at higher doping
concentrations. A spin feedback effect from the hopping term is
not included either, which results in a shift of $\lambda $ to
$\lambda _{m}$ in $E_{m}$ \cite{phasestring_meanfield}, without
qualitatively changing the physical consequences.

\subsection{Spin dynamics in superconducting ground state}

\label{sec:spin-dynamics-sc}

\begin{figure}[tbph]
\includegraphics{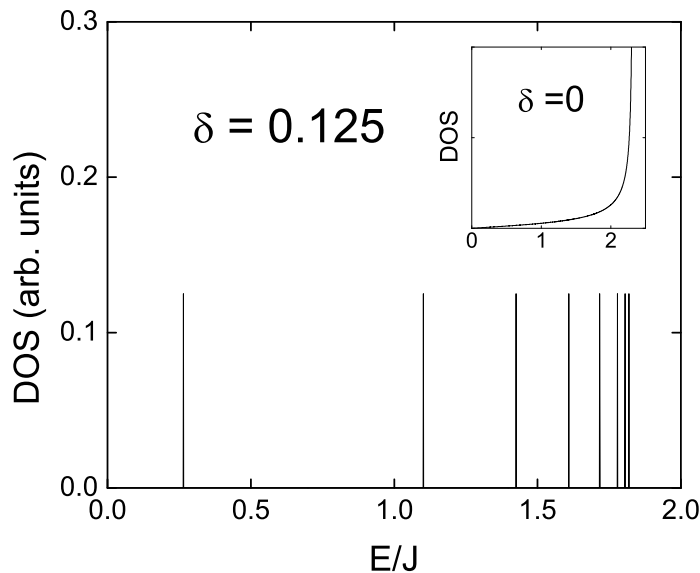}
\caption{The density of states (DOS) of the mean-field spinon spectrum $%
E_{m} $ at doping $\protect\delta =0.125$. Inset: the DOS in the AF state at
half filling. }
\label{fig:quasiparticleenergyspectrum} %
\centering
\end{figure}

\subsubsection{\protect\bigskip Excitation spectrum $E_{m}$}

According to the mean-field scheme outlined above, we can numerically
determine the mean-field `spinon' spectrum $E_{m}$ defined in (\ref{em}).

As an example, we solve the eigenequation (\ref{eq:3}) and self-consistent
equations (\ref{ds}) and (\ref{lambda}) at doping concentration $\delta
=0.125$. The chemical potential $\lambda $ is found to be $1.819J$ while the
RVB order parameter $\Delta ^{s}$ is $0.993$. In contrast, at half filling,
the results are $\lambda =2.316J$ and $\Delta _{0}^{s}=1.158$.

In Fig. \ref{fig:quasiparticleenergyspectrum}, the density of states (DOS)\
of the spectrum $E_{m}$ for $\delta =0.125$ is shown in the main panel,
while the half-filling case is plotted in the inset for comparison. The
figure shows that the two spectra are qualitatively very different. At
half-filling, the spectrum is continuous and gapless, with a large density
of states at the maximal energy which is slightly above $2J$. In the
superconducting state, the spectrum becomes discretized levels. This
discrete levels are due to the fact that the spectrum $\xi _{m}$ as the
solution of (\ref{eq:3}) has a Hofstadter spectrum as the result of the
vector potential $A_{ij}^{h}$ given in (\ref{uflux}). Note that the
distribution of the Landau-level-like structure in Fig. \ref%
{fig:quasiparticleenergyspectrum} remains uneven, which reflects the fact
that the average density of states increases with energy, as seen at
half-filling. The maximal energy is slightly less than $2J$ at $\delta
=0.125 $.

It is important to note that there is a gap between the lowest discrete
level and zero energy, which is $\sim 0.265J$ for $\delta =0.125$. There
will no more spinon Bose condensation $n_{BC}^{b}\neq 0$ such that the AFLRO
no longer exists.

\begin{figure}[tbph]
\centering
\includegraphics{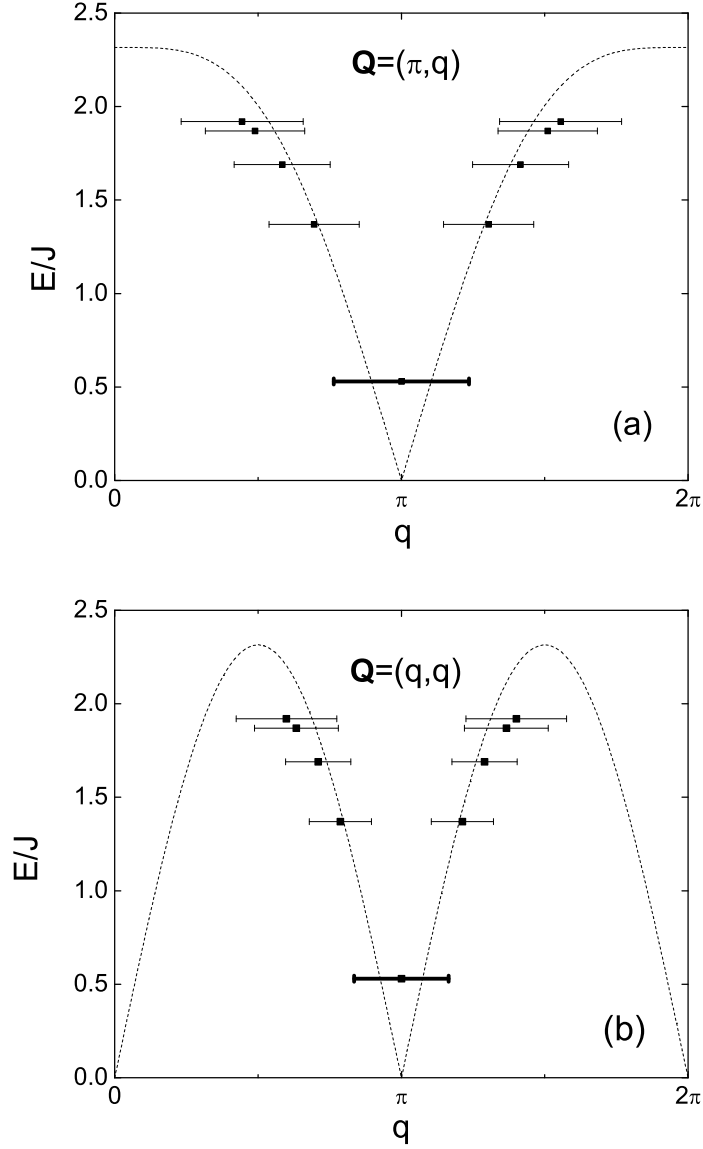}
\caption{The dispersive behavior of the spin excitation in the
superconducting state ($\protect\delta =0.125$), in comparison with the
spin-wave dispersion at half-filling (dashed curve). The peak positions of $%
\protect\chi ^{\prime \prime }$ in $\mathbf{Q}$- and $\protect\omega $-space
are shown along different $\mathbf{Q}$-directions: (a) along the diagonal
direction, $\mathbf{Q}=(q,q);$ (b) along $\mathbf{Q}=(\protect\pi ,q)$. The
solid bars mark the widths of the peaks in the momentum space (see text)$.$}
\label{fig:dispersion}
\end{figure}

\subsubsection{\protect\bigskip Dynamic spin susceptibility}

After diagonalizing the effective Hamiltonian $H_{s}$, the spin
susceptibility can be obtained straightforwardly. Due to the spin rotational
invariance \cite{phasestring_meanfield}, one may only consider the \^{z}%
-component susceptibility, which can be derived based on the Matsubara
Green's function $-\langle T_{\tau }S_{j}^{z}(\tau )S_{i}^{z}(0)\rangle $.
With the standard procedure outlined in Ref.\cite{phasestring_meanfield},
the imaginary part of the dynamic spin susceptibility at zero temperature is
given by
\begin{equation}
\chi ^{\prime \prime }(\mathbf{Q},\omega )=\frac{\pi }{8}\sum_{mm^{\prime
}}C_{mm^{\prime }}(\mathbf{Q})\left( \frac{\lambda ^{2}-\xi _{m}\xi
_{m^{\prime }}}{E_{m}E_{m^{\prime }}}-1\right) \mathrm{sgn}(\omega )\delta
(|\omega |-E_{m}-E_{m^{\prime }}),  \label{eq:6}
\end{equation}%
where
\begin{equation}
C_{mm^{\prime }}(\mathbf{Q})\equiv \frac{1}{N}\sum_{ij}e^{i\mathbf{Q}\cdot (%
\mathbf{x}_{i}-\mathbf{x}_{j})}\omega _{m\sigma }(i)\omega _{m\sigma }^{\ast
}(j)\omega _{m^{\prime }\sigma }^{\ast }(i)\omega _{m^{\prime }\sigma }(j).
\label{C}
\end{equation}

The discrete energy levels of $E_{m}$ illustrated in Fig.\ref%
{fig:quasiparticleenergyspectrum} will show up in $\chi ^{\prime \prime }(%
\mathbf{Q},\omega ).$ We plot the positions of these peaks in $\chi ^{\prime
\prime }(\mathbf{Q},\omega )$ in energy and momentum space, as well as the
FWHM (full width of half maximum) in momentum space, in Fig. \ref%
{fig:dispersion}. The momentum scan in Fig. \ref{fig:dispersion}(a) is along
$(\pi ,q)$ direction and is along the diagonal $(q,q)$ in Fig. \ref%
{fig:dispersion}(b). One sees that each discrete energy corresponds to a
finite width in momentum as depicted by a finite bar.

For comparison, the spin-wave peak positions at half-filling are shown as
dotted curves in Fig.\ref{fig:dispersion}. At $\delta =0.125,$ although the
spin excitations are no longer propagating modes, as evidenced by the flat
(dispersionless) small bars at discrete energies, the envelope of the
overall spectrum at high energies still approximately track the dispersion
of the spin wave at half-filling, with a slightly softened spin-wave
velocity. Note that there are actually some more peaks at even higher
energies than in Fig.\ref{fig:dispersion}, but their weight is much reduced
due to the coherent factors in $\chi ^{\prime \prime }(\mathbf{Q},\omega )$
(see the local spin susceptibility below).

Fig.\ref{fig:dispersion} clearly depicts how the spin excitations in the
superconducting state continuously evolves from the spin-wave picture at
half-filling. The remnant high-energy spin wave signature at finite doping
is a very unique feature in this approach. Recently, such a high-energy spin
wave feature has been reported \cite{buyers} in underdoped \textrm{YBa}$_{2}%
\mathrm{Cu}_{3}\mathrm{O}_{6.5}$ compound.

In the following, we turn our attention to the lowest peak in Fig.\ref%
{fig:dispersion}, which has the largest weight as marked by the darkest FWHM
bar.

\begin{figure}[tbph]
\centering  \includegraphics{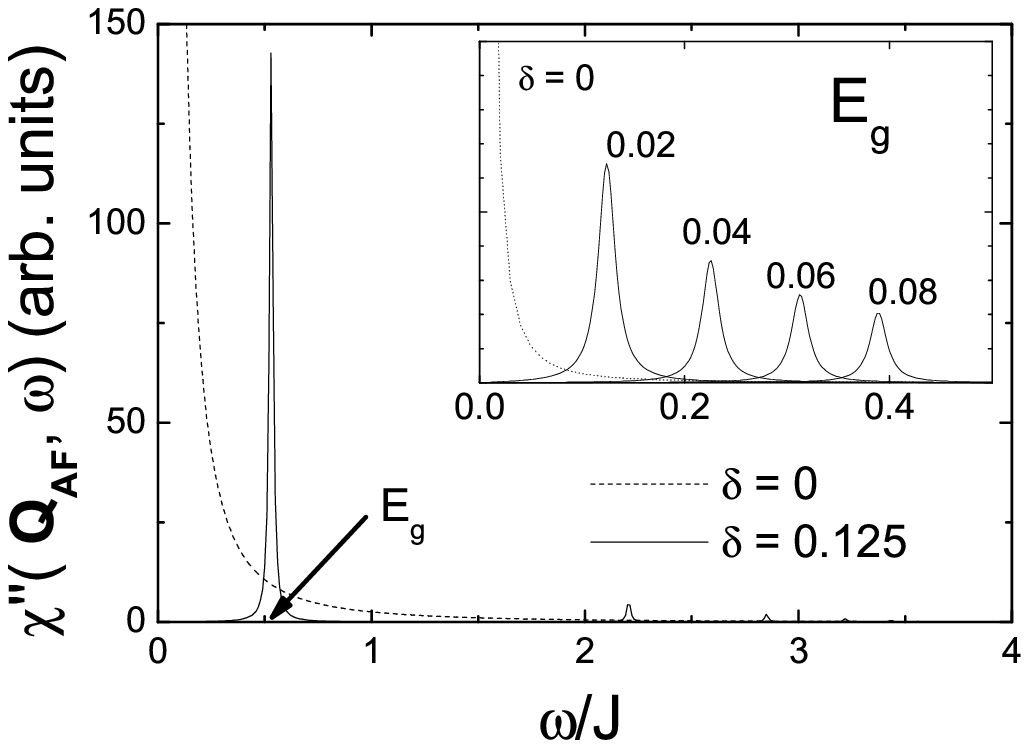}
\caption{Dynamic spin susceptibility $\protect\chi ^{\prime \prime }(\mathbf{%
Q}_{\mathrm{AF}}=(\protect\pi ,\protect\pi ),\protect\omega )$ in the
superconducting phase with $\protect\delta =0.125$ (solid curve). $E_{g}$
denotes the position of the resonancelike peak. The dotted curve is for the
AF state at half filling. Inset: the evolution of the resonance peak at
various dopings.}
\label{fig:pipi}
\end{figure}

\begin{figure}[tbph]
\centering  \includegraphics{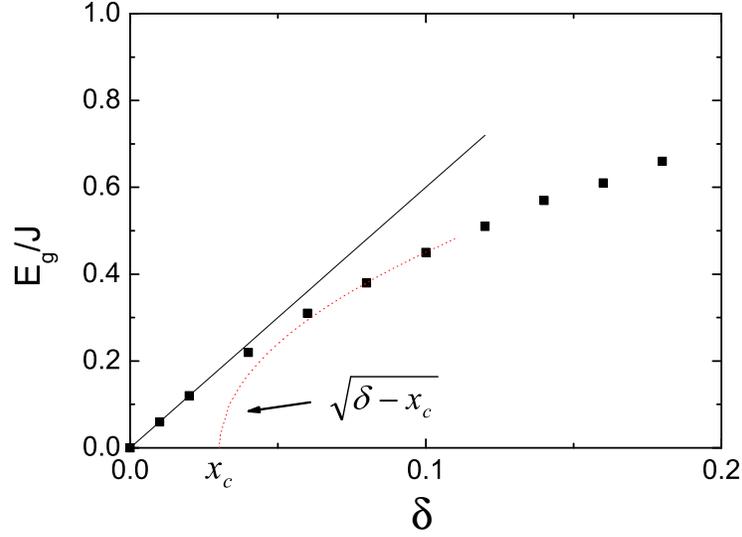}
\caption{The doping dependence of the resonancelike peak energy $E_{g}$. The
straight line illustrates the linear doping dependence at small $\protect%
\delta $. The dashed curve shows a $E_{g}\propto \protect\sqrt{\protect%
\delta -x_{c}}$ behavior if the AF state survives at a finite doping $x_{c},$
as shown in Ref. \protect\cite{kou}.}
\label{fig:res_vs_doping}
\end{figure}

\subsubsection{\protect\bigskip Resonancelike peak around AF wave-vector $%
\mathbf{Q}_{\mathrm{AF}}$}

\label{sec:prot-reson-peak}

Let us consider two special momenta, $\mathbf{Q}_{0}=(0,0)$ and $\mathbf{Q}_{%
\mathrm{AF}}=(\pi ,\pi )$. For $\mathbf{Q}=\mathbf{Q}_{0}$, with the
relation $\sum_{i}\omega _{m\sigma }(i)\omega _{m^{\prime }\sigma }^{\ast
}(i)=\delta _{mm^{\prime }}$, we have $C_{mm^{\prime }}(\mathbf{Q}_{0})=%
\frac{1}{N}\delta _{mm^{\prime }}$ such that
\begin{equation*}
\chi ^{\prime \prime }(\mathbf{Q}_{0},\omega )=\frac{\pi }{8N}\sum_{m}\left(
\frac{\lambda ^{2}-\xi _{m}^{2}-E_{m}^{2}}{E_{m}^{2}}\right) \mathrm{sgn}%
~(\omega )\delta (|\omega |-2E_{m})=0.
\end{equation*}%
Namely, there is no signature of $\chi ^{\prime \prime }(\mathbf{Q},\omega )$
at the ferromagnetic momentum $\mathbf{Q}_{0}.$

At the AF momentum $\mathbf{Q}_{\mathrm{AF}}$, one has
\begin{equation}
C_{mm^{\prime }}(\mathbf{Q}_{\mathrm{AF}})=\frac{1}{N}\sum_{ij}(-1)^{i-j}%
\omega _{m\sigma }(i)\omega _{m\sigma }^{\ast }(j)\omega _{m^{\prime }\sigma
}^{\ast }(i)\omega _{m^{\prime }\sigma }(j).  \label{eq:5}
\end{equation}%
In the eigenequation \eqref{eq:3}, it can be easily shown that for any given
state $m$ there is a corresponding state $\bar{m}$ with the relation $\xi
_{m}=-\xi _{\bar{m}}$ and $\omega _{m\sigma }(i)=(-1)^{i}\omega _{\bar{m}%
\sigma }(i)$. Then \eqref{eq:5} is reduced to
\begin{equation*}
C_{mm^{\prime }}(\mathbf{Q}_{\mathrm{AF}})=\frac{1}{N}\sum_{ij}\omega
_{m\sigma }(i)\omega _{m\sigma }^{\ast }(j)\omega _{\bar{m}^{\prime }\sigma
}^{\ast }(i)\omega _{\bar{m}^{\prime }\sigma }(j)=\frac{1}{N}\delta _{m\bar{m%
}^{\prime }}.
\end{equation*}%
and the dynamic spin susceptibility at $\mathbf{Q}_{\mathrm{AF}}$ can be
simplified to
\begin{equation}
\chi ^{\prime \prime }(\mathbf{Q}_{\mathrm{AF}},\omega )=\frac{\pi }{4N}%
\sum_{m}\left( \frac{\xi _{m}^{2}}{E_{m}^{2}}\right) \mathrm{sgn}(\omega
)~\delta (|\omega |-2E_{m}).  \label{kaipi}
\end{equation}%
\ \

The numerical result of $\chi ^{\prime \prime }(\mathbf{Q}_{\mathrm{AF}%
},\omega )$ at $\delta =0.125$ is shown in Fig. \ref{fig:pipi} by the solid
curve. The dotted curve is calculated at half-filling, which diverges as $1/{%
\omega ^{2}}$ at $\omega \rightarrow 0$, in consistency with the spin-wave
theory. Thus, in the superconducting phase, a resonancelike peak appears at $%
\mathbf{Q}_{\mathrm{AF}}$ with a finite energy $E_{g}=0.53J$ at $0.125$
(twice bigger than that of $E_{m}$ shown in Fig.\ref%
{fig:quasiparticleenergyspectrum})$.$ Note that higher energy (harmonic)
peaks in $\chi ^{\prime \prime }(\mathbf{Q}_{\mathrm{AF}},\omega )$ are
greatly reduced in strength in Fig.\ref{fig:pipi} due to the coherence
factor $\frac{\xi _{m}^{2}}{E_{m}^{2}}$ in (\ref{kaipi}). So only the lowest
peak at $E_{g}$ is clearly exhibited around $\mathbf{Q}_{\mathrm{AF}}$.

We further plot the resonancelike peak energy $E_{g}$ as a function of hole
concentration in Fig. \ref{fig:res_vs_doping}. At small doping, $E_{g}$ is
linearly proportional to $\delta ,$ $E_{g}=3.3\delta J,$ which is
extrapolated to zero at half-filling, where the gapless spin wave is
recovered. Note that in the present approach, the superconducting ground
state is extrapolated to $\delta =0^{+}$. In a more careful study of the
low-doping regime (beyond the mean-field approximation in the phase string
model) has revealed that the AF state actually will survive up to a finite
doping concentration, $\delta <x_{c}\simeq 0.043$ \cite{kou}. In that case,
one finds \cite{kou} that $E_{g}$ vanishes at $\delta =x_{c}$ following a
square root behavir: $E_{g}\propto \sqrt{{\delta -x_{c}}}$ as $\delta
\rightarrow x_{c}$, as shown by the dashed curve in Fig. \ref%
{fig:res_vs_doping}.

\begin{figure}[tbph]
\centering  \includegraphics{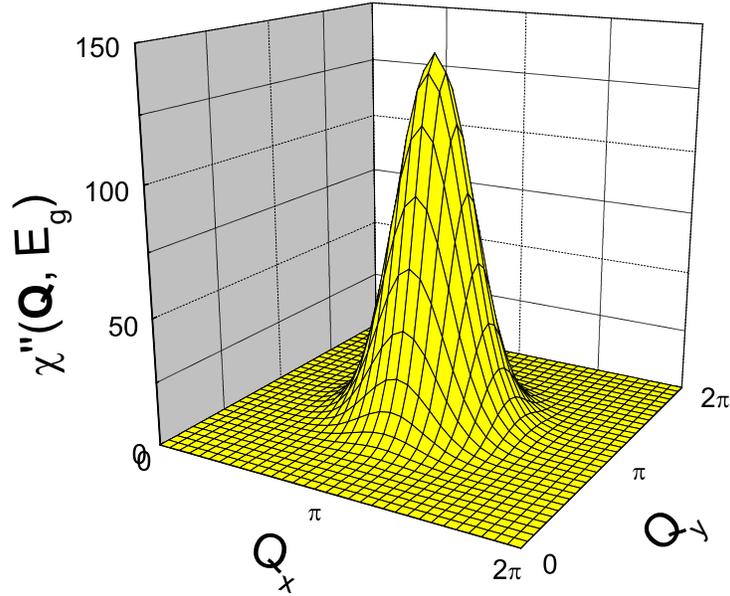}
\caption{Momentum distribution of $\protect\chi ^{\prime \prime }(\mathbf{Q},%
\protect\omega )$ at $\protect\omega =E_{g}$ ($\protect\delta =0.125)$.}
\label{fig:qprofile}
\end{figure}

\begin{figure}[tbph]
\centering  \includegraphics{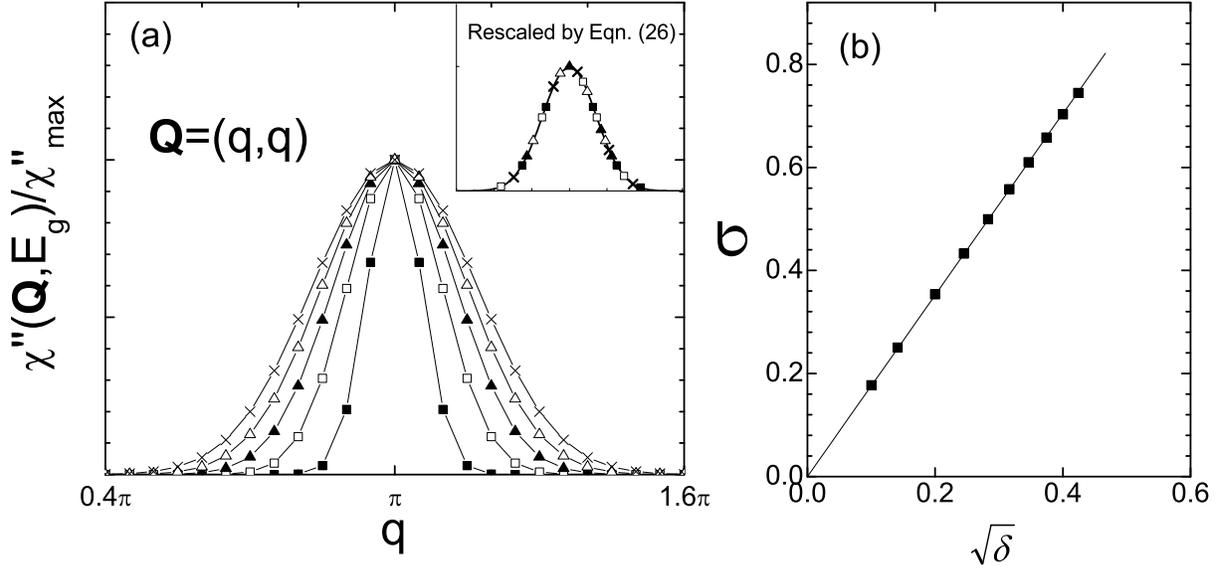}
\caption{(a) Momentum distribution of $\protect\chi ^{\prime \prime }(%
\mathbf{Q},E_{g})$, scanned along the diagonal direction $\mathbf{Q}=(q,q)$
at various hole concentrations. The intensities are normalized at the
maximums. The inset shows that the data in the main panel can be well fit
into a Gaussian function $\exp (-(\mathbf{Q}-\mathbf{Q}_{\mathrm{AF}})^{2}/2%
\protect\sigma ^{2})$, with $\protect\sigma $ being scaled as linearly
proportional to$\protect\sqrt{\protect\delta }$, as shown in (b).}
\label{fig:qwidth}
\end{figure}

The momentum profile of the resonancelike peak at $E_{g}$ is shown in a
three-dimensional plot in Fig. \ref{fig:qprofile} at $\delta =0.125$. It
shows an intrinsic broadening of $\chi ^{\prime \prime }(\mathbf{Q},E_{g})$
in momentum around $\mathbf{Q}_{\mathrm{AF}}$, which can be well fit by a
Gaussian distribution function
\begin{equation}
\chi ^{\prime \prime }(\mathbf{Q},E_{g})\propto \exp \left( -\frac{(\mathbf{Q%
}-\mathbf{Q}_{\mathrm{AF}})^{2}}{2\sigma ^{2}}\right) .  \label{qshape}
\end{equation}%
The results for different hole concentrations are given in Fig. \ref%
{fig:qwidth}(a) along the diagonal momenta $\mathbf{Q}=(q,q).$ One can
adjust $\sigma $ to make all data well collapse onto a single Gaussian
function of (\ref{qshape}) as shown in the inset of Fig. \ref{fig:qwidth}%
(a). The obtained broadening $\sigma $ turns out to be nicely scaled
linearly with$\sqrt{\delta }$ [see in Fig. \ref{fig:qwidth}(b)]. Similar
plots can be done along different $\mathbf{Q}$ scans centered at $\mathbf{Q}%
_{AF}$ and generally one has $\sigma \propto \sqrt{\delta }$ in all
directions.

\begin{figure}[tbph]
\centering  \includegraphics{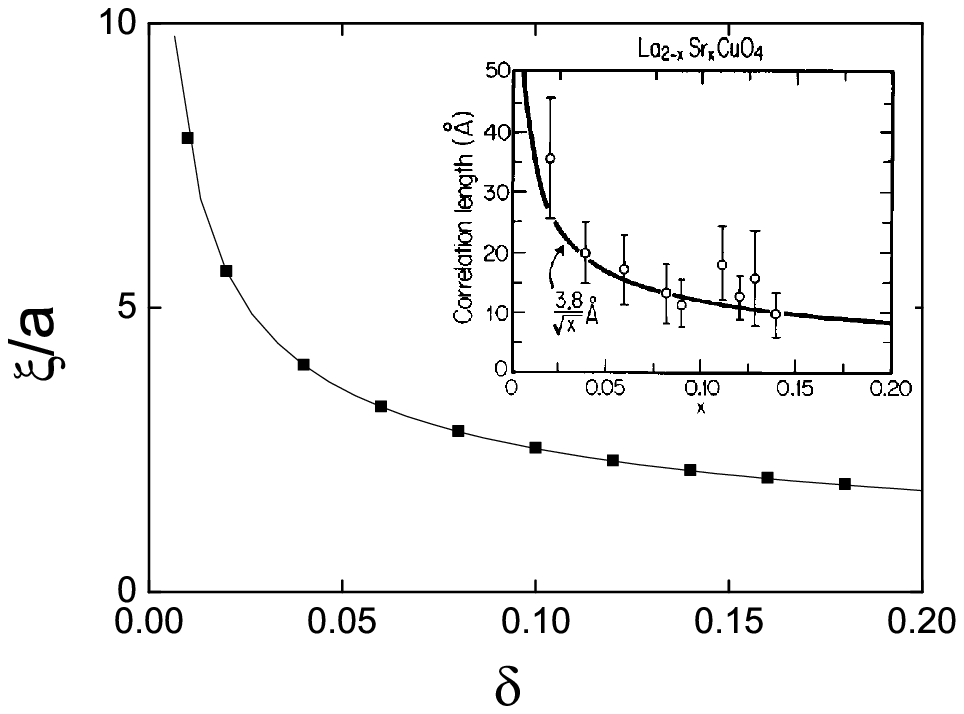}
\caption{The relation between the spin correlation length $\protect\xi /a$
and the hole concentration $\protect\delta $. The solid curve is $\frac{2}{%
\protect\pi \protect\sqrt{\protect\delta }}$. The inset is the experimental
results given in Ref. \protect\cite{LSCO}.}
\label{fig:correlationlength}
\end{figure}

If we neglect the small anisotropy along different momentum directions
centered at $\mathbf{Q}_{\mathrm{AF}}$ and perform a Fourier transformation
to (\ref{qshape}), we obtain the real-space correlation
\begin{equation}
\chi ^{\prime \prime }(\mathbf{R},\omega )\propto \exp (-\frac{\sigma ^{2}%
\mathbf{R}^{2}}{2})\equiv \exp (-\frac{\mathbf{R}^{2}}{\xi ^{2}})
\end{equation}%
with $\xi =\frac{\sqrt{2}}{\sigma }$. Thus, the spin-spin correlation
function decays exponentially with the distance in the superconducting
phase. This is consistent with a spin gap $E_{g}$ opening up in the spin
excitation spectrum. In Fig. \ref{fig:correlationlength}, $\xi $ is well fit
by the solid curve%
\begin{equation}
\xi =a\sqrt{\frac{2}{\pi \delta }.}  \label{xi0}
\end{equation}%
In the inset, the experimental result obtained in LSCO \cite{LSCO} is
presented for comparison. The general trend of $\xi /a\propto 1/\sqrt{\delta
}$ in both the experiment and theory is quite telling.

In Sec. III, we shall further discuss the momentum profile and longer spin
correlation lengths at \emph{lower} energies,, related to those seen in the
LSCO compound \cite{LSCO,yamada}, when the fluctuation effect is considered.

\begin{figure}[tbph]
\centering  \includegraphics{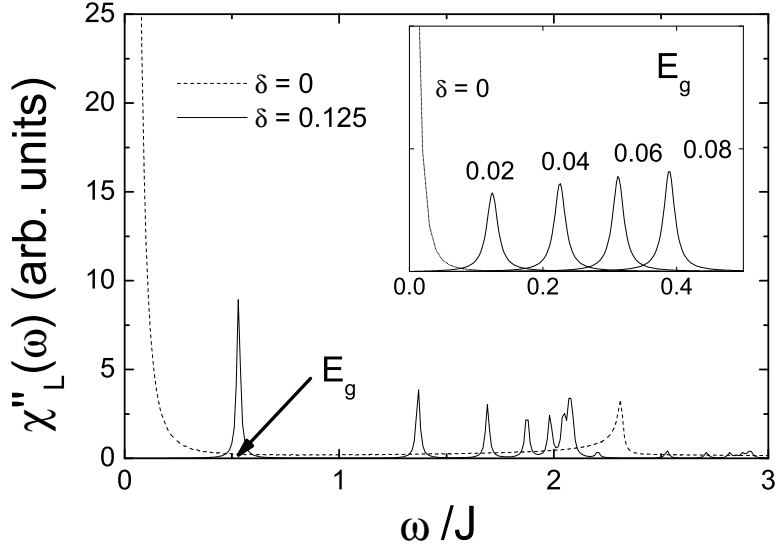}
\caption{Local susceptibility $\protect\chi _{L}^{\prime \prime }(\protect%
\omega )$ in the superconducting phase (solid curve) with $\protect\delta %
=0.125$, and at half filling (dashed curve). Inset: the evolution of the
lowest peak (the resonance peak) at different dopings.}
\label{fig:local}
\end{figure}

\subsubsection{\protect\bigskip Local susceptibility and spin sum rule}

The local spin susceptibility $\chi _{L}^{\prime \prime }(\omega )$ is also
an important quantity. It is defined by
\begin{equation*}
\chi _{L}^{\prime \prime }(\omega )=\int \frac{d^{2}\mathbf{Q}}{\left( 2\pi
\right) ^{2}}\chi ^{\prime \prime }(\mathbf{Q},\omega ),
\end{equation*}%
which describes the on-site spin-spin correlation. Based on \eqref{eq:6},
one obtains
\begin{equation}
\chi _{L}^{\prime \prime }(\omega )=\frac{\pi }{8}\sum_{mm^{\prime
}}K_{mm^{\prime }}\left( \frac{\lambda ^{2}}{E_{m}E_{m^{\prime }}}-1\right)
\mathrm{sgn}(\omega )\delta (|\omega |-E_{m}-E_{m^{\prime }}),
\end{equation}%
where
\begin{equation}
K_{mm^{\prime }}\equiv \frac{1}{N}\sum_{i}|\omega _{m\sigma }(i)|^{2}|\omega
_{m^{\prime }\sigma }(i)|^{2}.
\end{equation}

The numerical results of $\chi _{L}^{\prime \prime }(\omega )$ at $\delta
=0.125$ and $\delta =0$ are presented in Fig.\ref{fig:local} by the solid
and dashed curves, respectively. The low-energy parts in both cases are
similar to those seen in $\chi ^{\prime \prime }(\mathbf{Q},\omega )$ around
$\mathbf{Q}_{\mathrm{AF}}$ (Fig. \ref{fig:pipi}), as the AF correlations are
dominant at low energies. At high energies, more excitations which in
momentum space disperse away from $\mathbf{Q}_{\mathrm{AF}},$ as shown in
Fig.\ref{fig:dispersion}, are clearly present in $\chi _{L}^{\prime \prime
}(\omega ).$ We see that the main band extends up to $\sim 2.3J$ at
half-filling, while is slightly reduced to around $\sim 2.1J$ at $\delta
=0.125.$ These upper-bound spin excitations are expected to be seen near the
Brillouin zone boundary (see Fig.\ref{fig:dispersion}).

\begin{figure}[tbph]
\centering  \includegraphics{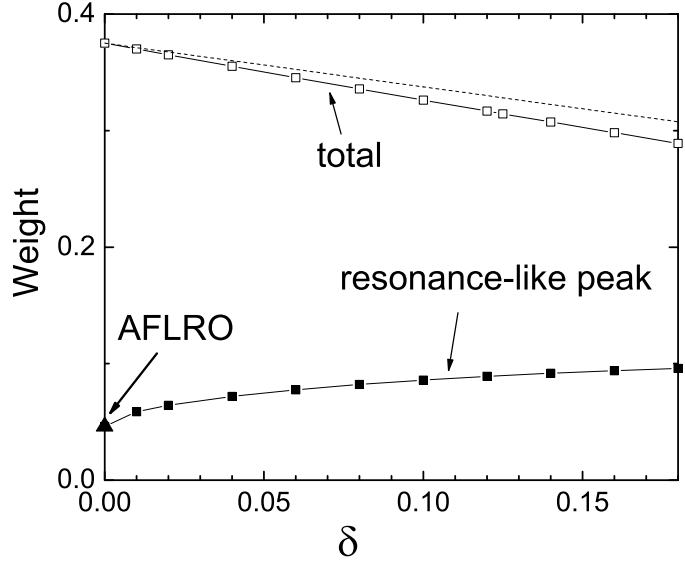}
\caption{The spin spectral weight vs doping. Open squares: the total weight;
closed squares: the weight of the resonancelike peak around $\mathbf{Q}_{%
\mathrm{AF}}$; closed triangles: the weight of the AFLRO peak at half
filling. The dashed line is the total weight from the exact sum rule, which
is rescaled to coincide with the mean-field value at $\protect\delta =0$ in
order to compare the doping dependence$.$}
\label{fig:weight}
\end{figure}

Although the intensity of each peak is physically not very meaningful, the
weigh of the peak is. The reason is that there is a sum rule about the local
dynamic spin susceptibility:
\begin{equation}
W_{\mathrm{total}}\equiv \int \mathrm{d}\omega ~\left[ 1+n(\omega )\right]
\chi _{L}^{\prime \prime }(\omega )=\langle (S_{i}^{z})^{2}\rangle ,
\end{equation}%
where the Bose distribution $n(\omega )=1/\left( e^{\beta \omega }-1\right)
. $ It means that the total weight of the spin susceptibility is related to
an averaged spin number per site. At half filling, it is obviously that $%
\langle (S_{i}^{z})^{2}\rangle $ is exactly $1/4$. At finite doping, $%
\langle (S_{i}^{z})^{2}\rangle $ should be reduced $(1-\delta )/4$.

In the bosonic RVB mean field state, the total weight can be calculated by
\begin{equation}
W_{\mathrm{total}}=\frac{1}{4N}\sum_{i\sigma }\langle b_{i\sigma }^{\dagger
}b_{i\sigma }\rangle \left( 1+\langle b_{i\sigma }^{\dagger }b_{i\sigma
}\rangle \right) .
\end{equation}%
By using $\langle b_{i\sigma }^{\dagger }b_{i\sigma }\rangle =(1-\delta )/2$%
, we have $W_{\mathrm{total}}=\frac{1}{8}(1-\delta )(3-\delta ).$ At
half-filling, the total weight is $\frac{3}{8}$ as compared to the exact
result $1/4$. The discrepancy is due to the relaxation of the no double
occupancy to a global level in the Schwinger-boson mean-field theory \cite%
{schwinger_boson}. In Fig. \ref{fig:weight}, the doping dependence of $W_{%
\mathrm{total}}$ is shown with the exact result (dashed line) rescaled at $%
\delta =0.$

We also show the integrated weight of the resonancelike peak in Fig. \ref%
{fig:weight} (solid curves with full squares), defined by
\begin{equation}
W_{\mathrm{peak}}\equiv \int_{\mathrm{peak}}\mathrm{d}\omega \left[
1+n(\omega )\right] ~~\chi _{L}^{\prime \prime }(\omega ).
\end{equation}%
At $\delta =0.125$, the weight of the peak is about $0.09,$ while the total
weight is about $0.314$, \emph{i.e.,} nearly $1/3$ of the total weight is
concentrated on the resonancelike peak. In Fig. \ref{fig:weight}, one can
see that with the increase of doping concentration, $W_{\mathrm{peak}}$
actually gets slightly increased, whereas $W_{\mathrm{total}}$ is reduced.
Namely, the resonancelike peak in the superconducting phase will become even
more prominent approaching the optimal doping from the underdoping. On the
other hand, as the doping concentration is reduced to zero, $W_{\mathrm{peak}%
}$ does not simply vanish. Instead, it approaches to a finite value which
precisely coincides with the weight of the delta function at $\omega =0$ and
$\mathbf{Q}=\mathbf{Q}_{\mathrm{AF}}$ in the dynamic spin structure function
at half-filling, which represents the AFLRO. Earlier on, we have seen that
at $\delta \rightarrow 0$ both $E_{g}$ and the width $\sigma $ of the peak
in momentum space go to zero. So the resonancelike peak continuously crosses
over to the AFLRO at half-filling.

\section{Spin dynamics beyond mean-field approximation}

So far our discussions on spin dynamics have all been based on a
generalized mean-field theory, characterized by the RVB order
parameter $\Delta^s$ [(\ref{brvb})]. Such a mean-field theory
works quite well at half-filling over a wide range of temperature
($\sim J/k_B$) in describing various ranges of spin-spin
correlations. In particular, the nearest-neighbor (nn) spin
correlation is directly related to $\Delta^s$ by
\begin{equation}
\left\langle \mathbf{S}_{i}\cdot \mathbf{S}_{j}\right\rangle _{nn}=-\frac{3}{%
8}\left\vert \Delta ^{s}\right\vert. \label{ss}
\end{equation}
It thus provides an important justification for the doped case:
Since spin-spin correlations, especially short-ranged ones, should
not be `washed out' immediately by the holes at small doping, the
nn RVB pairing $\Delta^s$ and thus the present mean-field state
underpinned by the RVB order parameter is expected to persist over
a finite range of doping, so long as the spin correlation length
is no less than the nn distance (i.e., the lattice constant). In
general, the effective Hamiltonian (\ref{hs}) is only valid within
a low-doping regime of $\Delta^s\neq 0$ which defines a pseudogap
regime in the phase string model. Since a spin gap opens at finite
doping in this regime, as shown in the last section, the amplitude
fluctuation of the RVB parameter usually is not very important.

Furthermore, we note that even within such a pseudogap phase
characterized by a finite $\Delta^s$, the effective RVB
description is not a usual mean-field theory beyond the
half-filling. Generally speaking, the effective spinon Hamiltonian
(\ref{hs}) is a gauge model, in which the topological gauge field
$A_{ij}^{h}$ describes $\pi $ fluxoids bound to holes according to (\ref{ah}%
). Namely, this is not a spinon-only model and the hole-doping
effect enters the Hamiltonian via $A_{ij}^{h}$, which represents
the nontrivial frustration on the spin degrees of freedom from the
motion of holes. In the previous section, the effect of
$A_{ij}^{h}$ has been treated in a mean-field approximation. In
the following, we shall discuss how to go beyond this mean-field
level.

\label{sec:spin-dynamics-with}

\subsection{Fluctuations induced by the density fluctuations of holes}

To examine the effect of fluctuations in $A_{ij}^{h}$ on spin
dynamics beyond the mean-field approxiamtion, one has to first
deal with the hole density fluctuations. In the phase string
model, the hole degrees of freedom is also dependent
\cite{phasestring_meanfield} on the spin degrees of freedom. The
nature of such mutually entangled charge and spin degrees of
freedom is expected to make the theory quite nontrivial in a
general case.

In the superconducting phase, a uniform holon condensation \cite%
{phasestring_meanfield} makes the topological gauge field $A_{ij}^{h}$
simplified as it may be treated as describing a uniform flux, namely, $%
A_{ij}^{h}\approx \bar{A}_{ij}^{h}$, with $\bar{A}_{ij}^{h}$ determined by
\begin{equation}
\sum_{C}\bar{A}_{ij}^{h}=\pi \sum_{l\in C}\bar{n}_{l}^{h}=\pi \sum_{l\in
C}\delta  \label{uflux1}
\end{equation}%
for an arbitrary loop $C$ according to (\ref{ah})$.$ In the previous
section, we have found that the spin dynamics in the superconducting phase
is qualitatively modified by such $\bar{A}_{ij}^{h}$ as compared to the
AFLRO state at half-filling.

However, the ideal Bose-Einstein condensation in treating $A_{ij}^{h}$ as $%
\bar{A}_{ij}^{h}$ is only an approximate description of the holon
condensation in the superconducting phase. In reality, one can expect all
kinds of hole density fluctuations. The fluctuation of $A_{ij}^{h},$\emph{\
i.e.,} $\delta A_{ij}^{h}\equiv A_{ij}^{h}-$ $\bar{A}_{ij}^{h}$, will be
tied to the density fluctation of the holes according to (\ref{ah}) as
follows:%
\begin{equation}
\sum_{C}\delta A_{ij}^{h}=\pi \sum_{l\in C}\delta n_{l}^{h}=\pi \sum_{l\in
C}(n_{l}^{h}-\delta )  \label{dflux}
\end{equation}%
In the following, we shall examine the effect of $\delta A_{ij}^{h}$ on the
spin susceptibility previously obtained in the mean-field approximation.

Since holons are condensed in the superconduting phase, we may still
reasonably neglect, to leading order of approximation, the dynamic
fluctuations in $\delta A_{ij}^{h}$ and only focus on the static spatial
fluctuations. As a Bose condensate is compressible, impurities and lattice
distortions can all lead to some microscopic spatial inhomogeneity of the
hole distribution, and below we introduce an approximate scheme to simulate $%
\delta A_{ij}^{h}$ related to a microscopically inhomogeneous distribution
of holes.
\begin{figure}[tbph]
\centering  \includegraphics{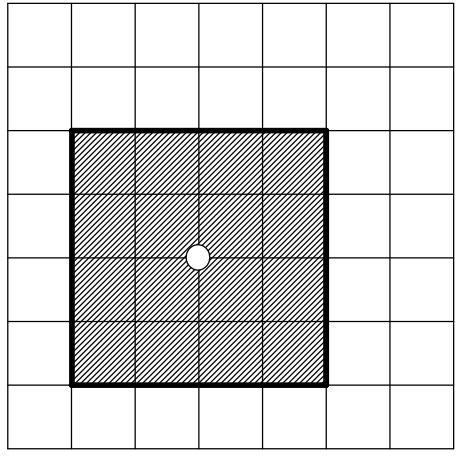} \caption{A way
to introduce the gauge-field fluctuations related to the holon
density. The open circle denotes a holon. The $\protect\pi
$-fluxoid bound to the holon is smeared to the shadow area which
is smaller than the whole lattice.} \label{fig:hole_smear}
\end{figure}

\begin{figure}[tbph]
\centering  \includegraphics{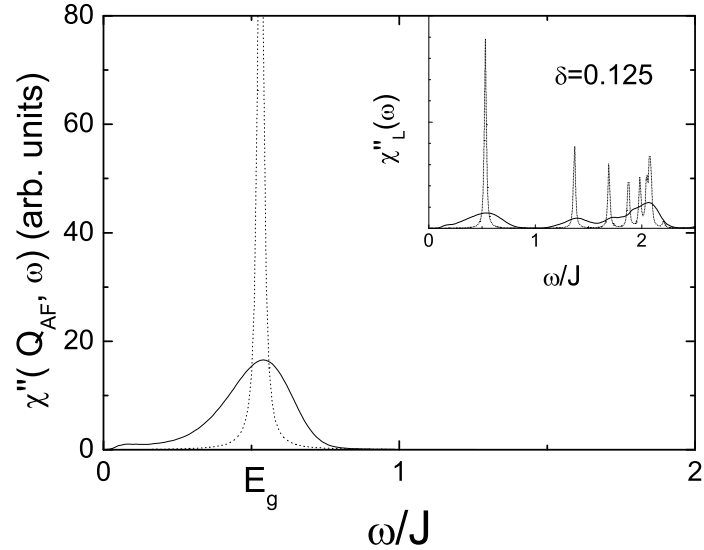}
\caption{$\protect\chi ^{\prime \prime }(\mathbf{Q}_{\mathrm{AF}},\protect%
\omega )$ with incorporating the fluctuations induced by the charge degrees
of freedom. The doping is at $0.125$, and the inset shows the local
susceptibility $\protect\chi _{L}^{\prime \prime }(\protect\omega )$ in the
same situation. }
\label{fig:pipi_with_fluc}
\end{figure}

We first smear each $\pi $ fluxoid bound to a hole within a finite size
(Fig. \ref{fig:hole_smear} shows one configuration), representing some
characteristic length scale of coherence for a bosonic holon, which should
be still much larger than the average hole-hole distance to reflect the
holon condensation. Then putting these smeared $\pi $ fluxoids randomly on
the lattice. If the smearing size of each $\pi $ fluxoid is infinite, then
the problem reduces back to the case of ideal Bose condensation with $\delta
A_{ij}^{h}=0$. For finite sizes of fluxoids, there generally exist intrinsic
fluctuations in the flux distribution of $A_{ij}^{h}$, which we use to
simulate the fluctuations related to the hole distribution$.$ Since it is
static, with each of such a configuration of non-uniform fluxes, we can
follow the steps in last section to get a non-uniform mean-field solution
and determine a dynamic spin susceptibility. The dynamic spin susceptibility
at $\mathbf{Q}_{\mathrm{AF}},$ averaged over the random configurations, is
presented in Fig. \ref{fig:pipi_with_fluc}, and the local susceptibility $%
\chi _{L}^{\prime \prime }(\omega )$ is shown in the inset. The result is
calculated in a $16\times 16$ lattice with each $\pi $ flux being smeared
within a $14\times 14$ lattice size, with more than 10,000 configurations
being averaged.

For comparison, the mean-field results are plotted as dashed curves in Fig. %
\ref{fig:pipi_with_fluc}. The main effect of such fluctuations in $A_{ij}^{h}
$ is to cause the broadening of the resonancelike peak as well as
high-energy peaks in energy space, although the peak positions, like $E_{g},$
essentially do not change. Since in the mean-field case, the discrete levels
are composed of degenerate Landau levels of $\xi _{m}$, a broadening due to
lifting up the degeneracies by the fluctuations in $A_{ij}^{h}$ can be
easily understood. So the above simple-minded approach to treat $\delta
A_{ij}^{h}$ provides some valuable insight into the fluctuation issue in the
framework of the bosonic RVB theory. A realistic treatment with a more
accurate profile of the lineshape in the dynamic spin susceptibility is
beyond the scope of this work.

\subsection{Incommensurability in momentum space}

\begin{figure}[bh]
\centering  \resizebox{120mm}{!}{%
\includegraphics{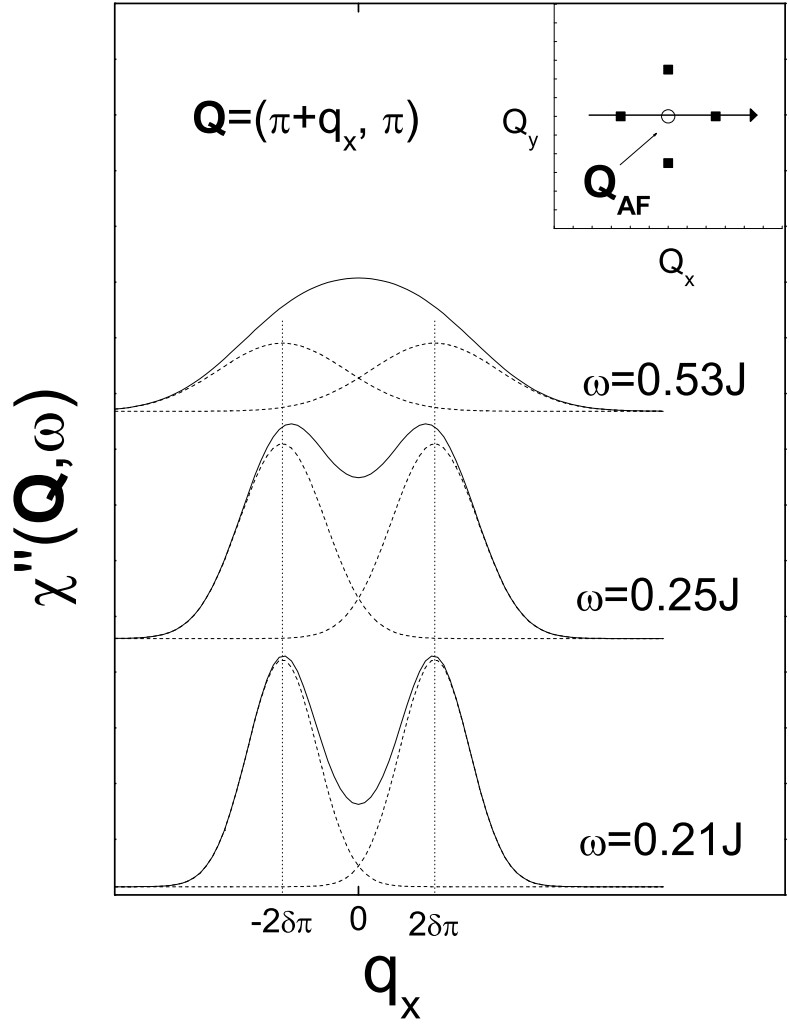}}
\caption{The incommensurate structure is generally presented in $\protect%
\chi ^{\prime \prime }(\mathbf{Q},\protect\omega )$ due to the
phase string effect \protect\cite{phasestring_inc}, but its
visibility depends on the
energy. The broad commensurate peak at $\protect\omega =E_{g}\sim 0.53J$ at $%
\protect\delta =0.125$ is actually composed of four peaks (dashed
curves), which becomes visibly `split' in momentum space as
$\protect\omega $ is lowered below $E_{g},$ when the fluctuational
effect is included, where the individual peak width is reduced (or
spin correlation length is enhanced) . }
\label{fig:incommensurability}
\end{figure}

The bosonic RVB mean-field state is based on the phase string formulation
\cite{phasestring_general} of the $t-J$ model, in which the short-distance
singular part of the phase string effect introduced by the hopping of holes
has been `gauged away' such that the Hamiltonian in the new formalism is
free of such singularities and thus becomes perturbatively treatable. But
when one considers the physical quantities like the dynamic spin
susceptibility, such singular effect should be still present in the
correlation function and has to be incorporated carefully. It has been shown
previously \cite{phasestring_inc} that the leading order contribution of
such a singular effect to the dynamic spin susceptibility is simply
represented by the incommensurate shifting of the momentum $\mathbf{Q}$ in $%
C_{mm^{\prime }}(\mathbf{Q)}$ defined in (\ref{C}) by $\delta Q_{x}\mathbf{%
=\pm }2\pi g$\textbf{\ }and $\delta Q_{y}=\mathbf{\pm }2\pi g$ $($taking $%
a=1)$ with $g\simeq \delta .$ However, since the momentum width of the
resonancelike peak in the mean-field is given by $\sigma =\sqrt{\pi \delta }%
, $ the incommensurability does not explicitly show up in the dynamic spin
susceptibility \cite{phasestring_inc} and the resonancelike peak still looks
like one peak centered at $\mathbf{Q}_{\mathrm{AF}}$, as illustrated in the
top panel of Fig. \ref{fig:incommensurability}.

Now, due to the above-discussed fluctuational effect, the resonancelike peak
is broadened with some of its weight shifting towards lower energies shown
in Fig. \ref{fig:pipi_with_fluc}. The corresponding width for these new
low-lying modes in momentum space will be reduced too (\emph{i.e., }the
spin-spin correlation lengths are enhanced at energies lower than $E_{g}$)
such that the incommensurability may become manifested in the dynamic spin
susceptibility gradually with the decrease of the energy. Indeed, by using
the same simulation used in Fig. \ref{fig:pipi_with_fluc}, the
incommensurate peaks do show up in the modified $\chi ^{\prime \prime }(%
\mathbf{Q},\omega )$ with incorporating the incommensurate shifting \cite%
{phasestring_inc}, as $\omega $ is lowered below $E_{g},$ which is
illustrated in Fig. \ref{fig:incommensurability} at $\delta =0.125$.

Therefore, the incommensurability in the dynamic spin susceptibility
function is an intrinsic property of the phase string effect \cite%
{phasestring_inc}. But its visibility crucially depends on spin fluctuations
with longer correlation lengths at low energies. Such low-lying spin
excitations, induced by the charge density fluctuations discussed above, are
usually most prominent in the single-layer case, applicable to the \textrm{%
LSCO }compound. In contrast, the charge density fluctuations are expected to
be weaker in the bilayer systems such as the \textrm{YBCO} compound, where
the interlayer coupling will prefer the uniform distribution of the holons
as to be discussed in the next section.

\section{Bosonic RVB description with interlayer coupling}

For a bilayer system, the $t-J$ model can be generalized as
\begin{eqnarray}
H_{t-J}^{\mathrm{bilayer}} &=&-t\sum_{\langle ij\rangle l\sigma }c_{il\sigma
}^{\dagger }c_{jl\sigma }-t_{\bot }\sum_{i\sigma }c_{i1\sigma }^{\dagger
}c_{i2\sigma }+H.c.  \notag \\
&&+J\sum_{\langle ij\rangle l}\mathbf{S}_{il}\mathbf{S}_{jl}+J_{\bot
}\sum_{i\sigma }\mathbf{S}_{i1}\cdot \mathbf{S}_{i2}  \label{bilayer}
\end{eqnarray}%
in which the additional subscript, $l=1,2,$ is the layer index. By
introducing an additional bosonic RVB order parameter
\begin{equation}
\Delta _{\bot }^{s}\equiv \sum_{\sigma }\langle b_{i1\sigma }b_{i2-\sigma
}\rangle ,  \label{eq:4}
\end{equation}%
and the Bogoliubov transformation
\begin{equation}
b_{il\sigma }=\sum_{mk}\omega _{mk\sigma }(i,l)(u_{mk}\gamma _{mk\sigma
}-v_{mk}\gamma _{mk-\sigma }^{\dagger })\text{ \ \ \ \ \ }
\end{equation}%
with $k=\pm ,$ the mean-field spinon Hamiltonian can be diagonalized in the
holon condensed phase in a procedure similar to Sec. II as given in Appendix
A. We find
\begin{eqnarray}
&&\omega _{mk\sigma }(i,l)=\frac{1}{\sqrt{2}}\left[ \mathrm{sgn}(\xi _{m})k%
\right] ^{l}\omega _{m\sigma }(i), \\
&&u_{mk}=\sqrt{\frac{1}{2}\left( \frac{\lambda }{E_{mk}}+1\right) }, \\
&&v_{mk}=\mathrm{sgn}(\xi _{mk})\sqrt{\frac{1}{2}\left( \frac{\lambda }{%
E_{mk}}-1\right) }.
\end{eqnarray}%
and the spinon energy spectrum
\begin{equation}
E_{mk}=\sqrt{\lambda ^{2}-\left\vert \xi _{mk}\right\vert ^{2}},
\label{eq:1}
\end{equation}%
in which
\begin{equation}
\xi _{mk}=\mathrm{sgn}(\xi _{m})\left( \left\vert \xi _{m}\right\vert +\frac{%
kJ_{\bot }\Delta _{\bot }^{s}}{2}\right) .  \label{eq:7}
\end{equation}%
In the above, $\xi _{m}$ and $\omega _{m\sigma }(i)$ are the solution of %
\eqref{eq:3}, as the counterparts of $\xi _{mk}$ and $\omega _{mk\sigma
}(i,l),$ respectively, in the single layer case. Finally, the
self-consistent equations of the RVB order parameters and the Lagrangian
multiplier $\lambda $ are given by
\begin{eqnarray}
\left\vert \Delta ^{s}\right\vert ^{2} &=&\frac{1}{4NJ}\sum_{mk}\frac{\xi
_{m}\xi _{mk}}{E_{mk}}\coth \frac{\beta E_{mk}}{2}, \\
\Delta _{\bot }^{s} &=&-\frac{1}{2N}\sum_{mk}\frac{\mathrm{sgn}(\xi
_{m})k\xi _{mk}}{E_{mk}}\coth \frac{\beta E_{mk}}{2}, \\
2-\delta &=&\frac{1}{2N}\sum_{mk}\frac{\lambda }{E_{mk}}\coth \frac{\beta
E_{mk}}{2}.
\end{eqnarray}

\begin{figure}[tbph]
\centering
\includegraphics{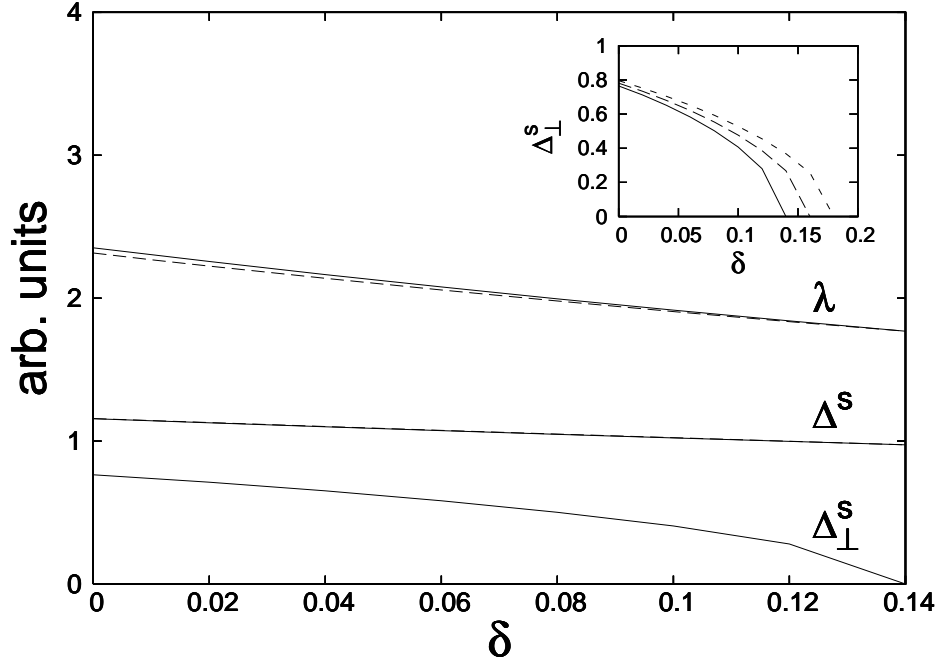}
\caption{
The doping dependence of $\Delta _{\bot }^{s}$, $\Delta _{s}$, $
\protect\lambda $. The solid line is at $J_{\bot }=0.1J$, the
dashed line is
the results in the single layer case. The inset is the doping dependence of $%
\Delta _{\bot }^{s}$ at different $J_{\bot }$'s: solid line, $J_{\bot
}=0.10J $; dashed line, $J_{\bot }=0.11J$, and dotted line, $J_{\bot }=0.12J$%
}
\label{fig:interlayer_parameter}
\end{figure}

Fig. \ref{fig:interlayer_parameter} shows the results obtained by the
self-consistent equations as functions of doping concentration (solid lines)
at $J_{\bot }=0.1J$, while the values of $\Delta ^{s}$ and $\lambda $ in the
single layer case are plotted by the dashed lines for comparison, which only
change slightly with the introduction of the interlayer coupling $J_{\bot
}=0.1J$.

In the inset of Fig. \ref{fig:interlayer_parameter}, the doping dependence
of $\Delta _{\bot }^{s}$ is shown at various $J_{\bot }$'s: $J_{\bot }=0.10J$%
, $0.11J,$ and $0.12J$. We note that $\Delta _{\bot }^{s}$ is comparable
with the interlayer pairing $\Delta ^{s}$ at half filling, \emph{e.g.,} $%
0.765$ versus $1.157$ even though $J_{\bot }=0.1J$ is quite small. This may
be attributed to the fact that the in-plane spin-spin correlation length $%
\xi $ is very large at half filling, which diverges at zero temperature. As
the consequence, spin mismatches between the two layers will involve a large
region determined by $\xi $, costing a big energy. This effectively enhances
the interlayer AF correlations and thus the interlayer RVB pairing $\Delta
_{\bot }^{s}$.

Away from half filling, as shown in Sec. \ref{sec:psmfa}, the in-plane
spin-spin correlation length decreases monotonically with the hole
concentration, which results in the reduction of the inflated interlayer AF\
correlations. Due to the competitive nature between $\Delta ^{s}$ and $%
\Delta _{\bot }^{s}$ (one spin cannot be part of two RVB pairs at the same
time), with the decrease of $\xi ,$ $\Delta _{\bot }^{s}$ will diminish much
faster than $\Delta ^{s},$ as shown in the main panel of Fig. \ref%
{fig:interlayer_parameter} as well as the inset for different $J_{\perp }$'s.

\subsection{Spinon spectrum:\ Bonding and antibonding states}

According to \eqref{eq:1} and \eqref{eq:7}, we find that with a finite $%
\Delta _{\bot }^{s}$, the original spinon spectrum $E_{m}$ in the
single-layer case is split into two branches, a bonding state $E_{m+}$ and
an antibonding state $E_{m-}$.

The DOS of spinon spectrum at half-filling is shown in the inset of Fig. \ref%
{fig:interlayer_spinon_spectrum}. At $\delta =0,$ $\xi _{m}$ reduces to $\xi
_{\mathbf{q}}=-J\Delta ^{s}(\cos q_{x}a+\cos q_{y}a)$ . The ground state
still has an AFLRO such that $\lambda =\max (\left| \xi _{\mathbf{q}%
,k}\right| )=(2J\Delta ^{s}+\frac{J_{\bot }\Delta _{\bot }^{s}}{2})$. The
DOS of the bonding states is the same as the single-layer case at $\omega
\rightarrow 0,$ while the antibonding states open a gap $=\min (E_{m-})=2%
\sqrt{J\Delta ^{s}}\sqrt{J_{\bot }\Delta _{\bot }^{s}}\sim 0.60J$ at $%
J_{\bot }=0.1J$ as shown by the dashed curve in the inset of Fig. \ref%
{fig:interlayer_spinon_spectrum}. This gap is approximately the same as the
gap in the dynamic spin susceptibility in the even channel (see below) as
observed by neutron scattering, which is about $70$ \textrm{meV} in
magnitude \cite{OpticalGap1, OpticalGap2}.

\begin{figure}[tbph]
\centering  \includegraphics{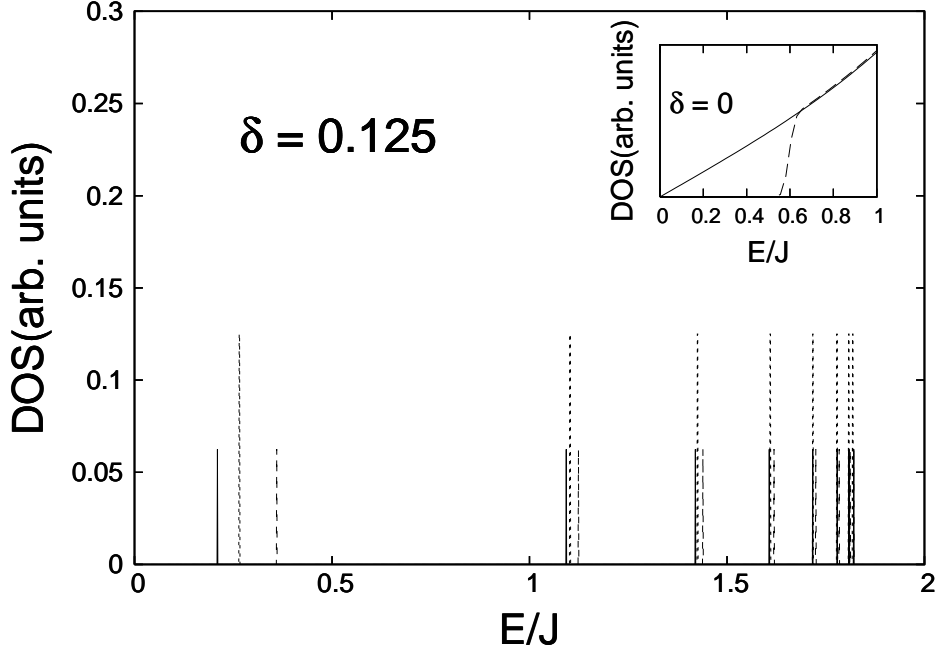} \caption{
The DOS of spinons at $\protect\delta =0.125$ in the bilayer case.
The solid line is for the bonding state, the dashed line for the
antibonding state, and the dotted line denotes the single layer
case for comparison. The inset shows the bilayer case at half
filling: solid curve is for the odd channel and dashed curve is
for the even channel. } \label{fig:interlayer_spinon_spectrum}
\end{figure}

The main panel of Fig. \ref{fig:interlayer_spinon_spectrum} shows the DOS of
spinon spectrum at $\delta =0.125$ and $J_{\bot }=0.1J$ in the
superconducting phase. Compared to the spinon spectrum in the single-layer
case with the Hofstadterlike structure illustrated by dotted lines in Fig. %
\ref{fig:interlayer_spinon_spectrum}, there are bilayer splittings between
the branches of the bonding (solid lines) and antibonding (dashed lines)
states, given by $E_{m-}-E_{m+}$. Because $J_{\bot }\Delta _{\bot }^{s}$ is
much smaller than $\lambda $, the splitting is most visible at the lowest
energy level where $\xi _{m}$ is the closest to $\lambda ,$ as shown in the
figure. In the following, we study how this bilayer splitting effect is
manifested in the dynamic spin susceptibility.

\subsection{Dynamic spin susceptibility}

In the bilayer case, the imaginary part of the spin susceptibility $\chi
^{\prime \prime }\left[ \left( \mathbf{Q},q_{\bot }\right) ,\omega \right] $
depends not only the in-plane wave vector $\mathbf{Q,}$ but also the c-axis
wave vector $q_{\bot }$. It can be shown that
\begin{equation}
\chi ^{\prime \prime }\left[ \left( \mathbf{Q},q_{\bot }\right) ,\omega %
\right] =\chi _{o}^{\prime \prime }(\mathbf{Q},\omega )\sin ^{2}(q_{\bot
}/2)+\chi _{e}^{\prime \prime }(\mathbf{Q},\omega )\cos ^{2}(q_{\bot }/2),
\end{equation}%
where $\chi _{o,e}^{^{\prime \prime }}$ is the imaginary part of the spin
susceptibility in the channels with odd and even symmetries, respectively,
obtained from the retarded versions of the Matsubara Green's functions
defined by
\begin{eqnarray}
\chi _{o}(i,j;\tau ) &=&-\langle T_{\tau }(S_{i1}^{z}(\tau )-S_{i2}^{z}(\tau
))(S_{j1}^{z}(0)-S_{j2}^{z}(0))\rangle , \\
\chi _{e}(i,j;\tau ) &=&-\langle T_{\tau }(S_{i1}^{z}(\tau )+S_{i2}^{z}(\tau
))(S_{j1}^{z}(0)+S_{j2}^{z}(0))\rangle .
\end{eqnarray}

With the same procedure as in Sec. \ref{sec:spin-dynamics-sc}, we can obtain
$\chi _{o}^{^{\prime \prime }}$ and $\chi _{e}^{^{\prime \prime }}$ at zero
temperature straightforwardly as follows
\begin{eqnarray}
\chi _{o}^{\prime \prime }(\mathbf{Q},\omega ) &=&\frac{\pi }{32}%
\sum_{mm^{\prime }kk^{\prime }}C_{mm^{\prime }}(\mathbf{Q})\left( 1-\mathrm{%
sgn}(\xi _{m}\xi _{m^{\prime }})kk^{\prime }\right) \left( \frac{\lambda
^{2}-\xi _{mk}\xi _{m^{\prime }k^{\prime }}}{E_{mk}E_{m^{\prime }k^{\prime }}%
}-1\right) \delta (\omega -E_{mk}-E_{m^{\prime }k^{\prime }}), \\
\chi _{e}^{\prime \prime }(\mathbf{Q},\omega ) &=&\frac{\pi }{32}%
\sum_{mm^{\prime }kk^{\prime }}C_{mm^{\prime }}(\mathbf{Q})\left( 1+\mathrm{%
sgn}(\xi _{m}\xi _{m^{\prime }})kk^{\prime }\right) \left( \frac{\lambda
^{2}-\xi _{mk}\xi _{m^{\prime }k^{\prime }}}{E_{mk}E_{m^{\prime }k^{\prime }}%
}-1\right) \delta (\omega -E_{mk}-E_{m^{\prime }k^{\prime }}).
\end{eqnarray}

\begin{figure}[tbph]
\centering
\resizebox{120mm}{!}{
  \begin{tabular}[t]{c}
    \includegraphics{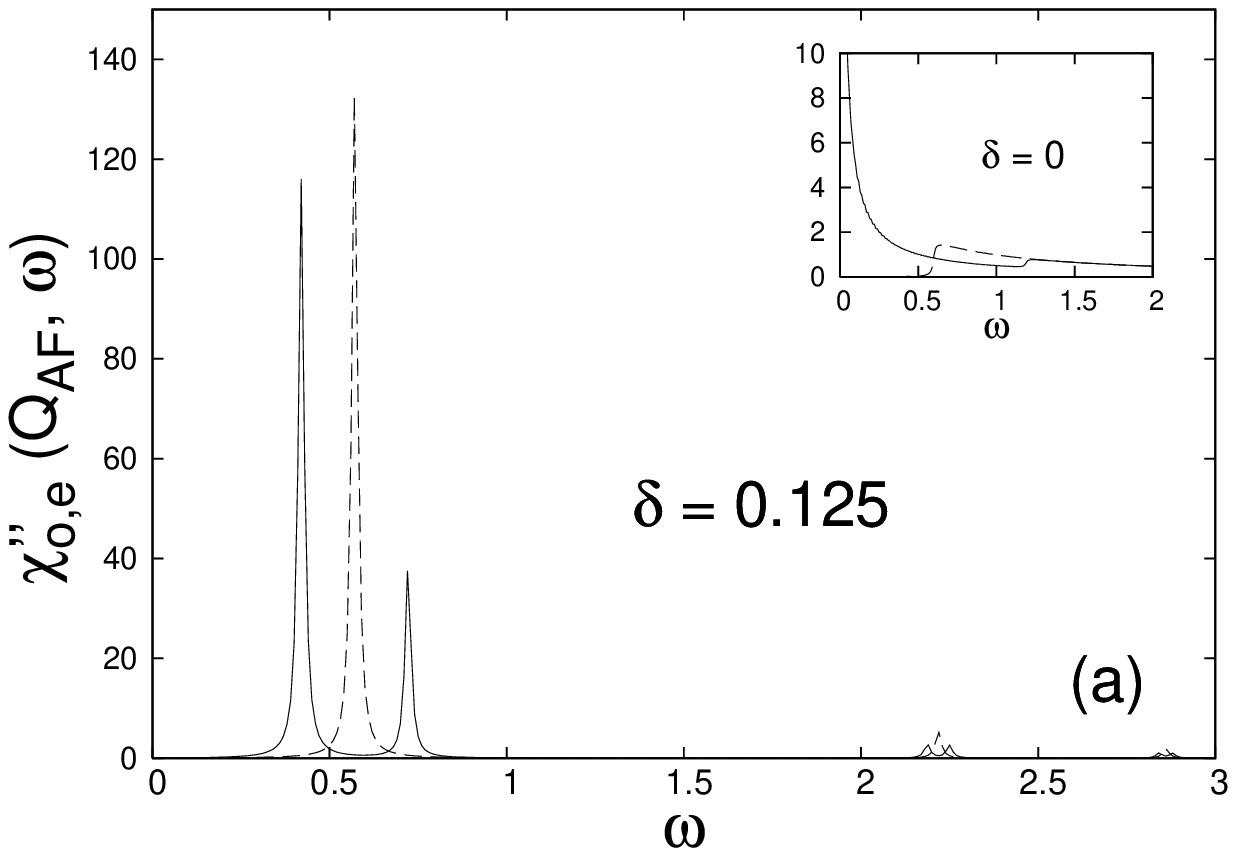} \\
    \includegraphics{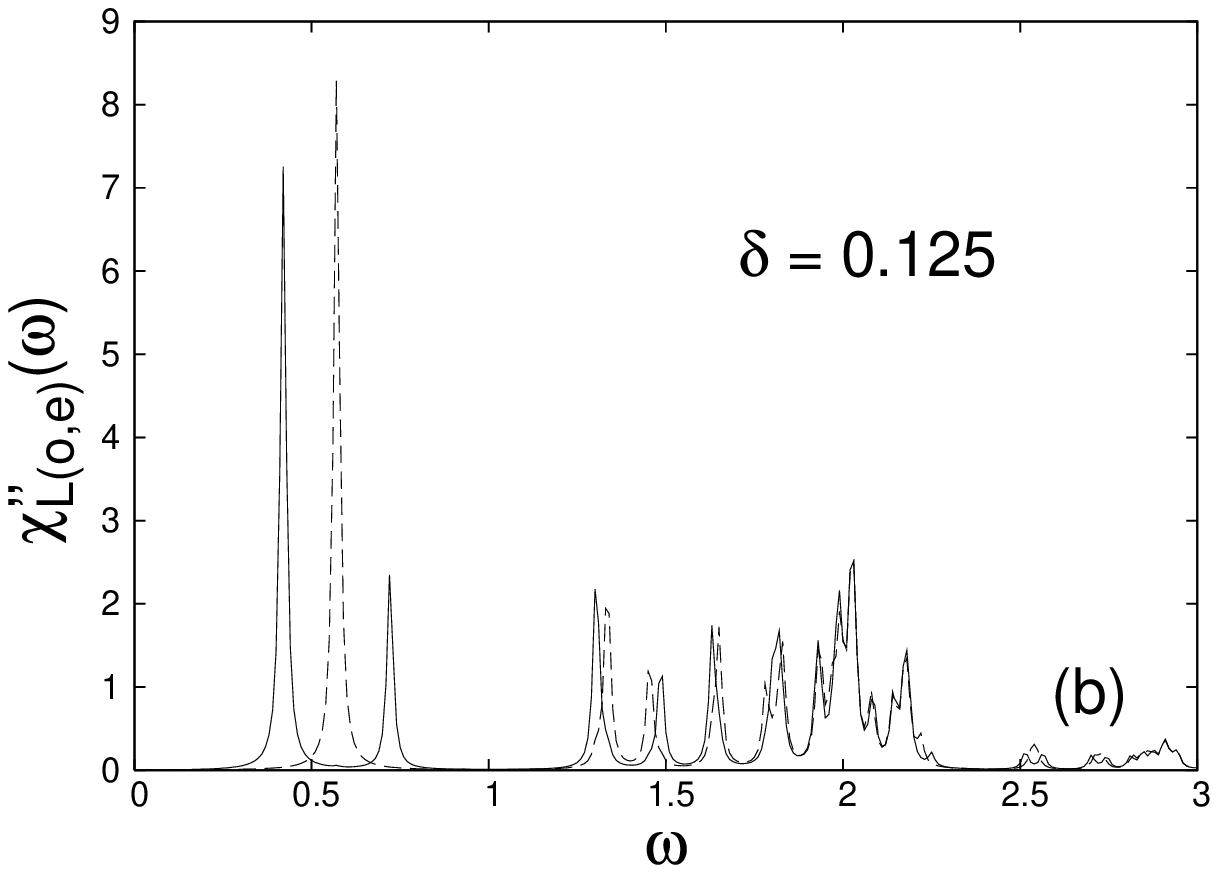}
  \end{tabular}}
\caption{(a) Dynamic spin susceptibility in the bilayer system at $\mathbf{Q}%
_{\mathrm{AF}}$ and $\protect\delta =0.125$. The solid line is for the odd
channel and the dashed line is for the even channel. The inset is for the
half filling case. (b) The local spin susceptibility. The solid curve is for
the odd channel, and the dashed curve is for the even channel.}
\label{fig:interlayer_pipi}
\end{figure}
According to Sec. \ref{sec:prot-reson-peak}, $C_{mm^{\prime }}(\mathbf{Q}_{%
\mathrm{AF}})=\frac{1}{N}\delta _{m\bar{m}^{\prime }}$ such that
\begin{eqnarray}
\chi _{o}^{\prime \prime }(\mathbf{Q}_{\mathrm{AF}},\omega ) &=&\frac{\pi }{%
8N}\sum_{mk}\frac{\xi _{mk}^{2}}{E_{mk}^{2}}\delta (\omega -2E_{mk}),
\label{eq:9} \\
\chi _{e}^{\prime \prime }(\mathbf{Q}_{\mathrm{AF}},\omega ) &=&\frac{\pi }{%
16N}\sum_{mk}\left( \frac{\lambda ^{2}+\xi _{mk}\xi _{m-k}}{E_{mk}E_{m-k}}%
-1\right) \delta (\omega -E_{mk}-E_{m-k}).  \label{eq:10}
\end{eqnarray}%
The above expressions clearly show that $\chi _{o}^{\prime \prime }(\mathbf{Q%
}_{AF},\omega )$ is solely contributed by a pair of spinon excitations both
from the bonding or antibonding states, while $\chi _{e}^{\prime \prime }(%
\mathbf{Q}_{AF},\omega )$ is contributed by a pair of spinon excitations,
one from the bonding state and the other from the antibonding state.
Compared to \eqref{kaipi}, one can see that $\chi _{o}^{\prime \prime }(%
\mathbf{Q}_{AF})$ is very similar to $\chi ^{\prime \prime }(\mathbf{Q}%
_{AF},\omega )$ in the single-layer case.

We present the numerical results at $J_{\bot }=0.1J$ and $\delta =0.125$ in
Fig. \ref{fig:interlayer_pipi}(a) and $\delta =0$ in the inset for
comparison. The solid curve represents the odd mode while the dashed curve
is for the even mode. From the main panel of Fig. \ref{fig:interlayer_pipi}%
(a), one sees that the single resonancelike peak in the single-layer case is
replaced by a double-peak structure corresponding to the lowest bonding and
antibonding states, respectively. In contrast, in the even channel, there is
only one peak whose center is just in the middle of the double peaks in the
odd channel. We also calculate the local spin susceptibility by integrating
over the in-plane wave vector $\mathbf{Q}$, which is given in Fig. \ref%
{fig:interlayer_pipi}(b). Two figures look quite similar.

\begin{figure}[tbph]
\centering  \includegraphics{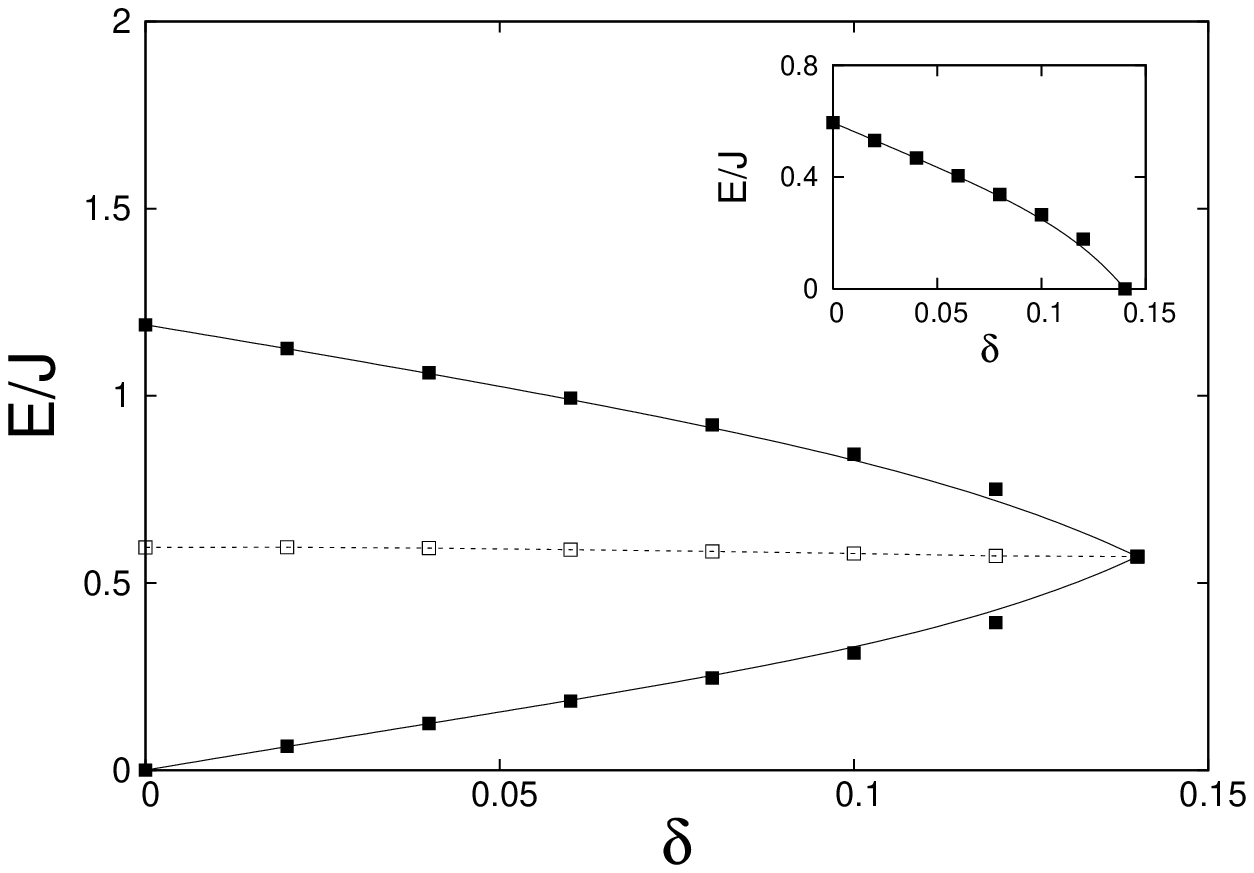}
\caption{ The energies of the peaks shown in Fig. \protect\ref%
{fig:interlayer_pipi}(a) as functions of doping. Close squares: the double
peaks in the odd channel; open square: the peak in the even channel. Inset:
the difference between the peak in the even channel and the lower energy one
in the odd channel.}
\label{fig:interlayer_peak}
\end{figure}

The doping dependences of the energies of these peaks are plotted in Fig. %
\ref{fig:interlayer_peak}, where the filled squares mark the double peaks in
the odd channel and the open squares describe the peak in the even channel.
One finds that the doping dependences for the three peaks are very
different. As $\delta $ tends to zero, the lowest peak in the odd channel
behaves like the resonancelike peak and reduces to the gapless spin wave
mode at half filling, while the peak at a higher energy in the same channel
move to high energy and reaches $1.190J$ finally.

\begin{figure}[tbph]
\centering  \includegraphics{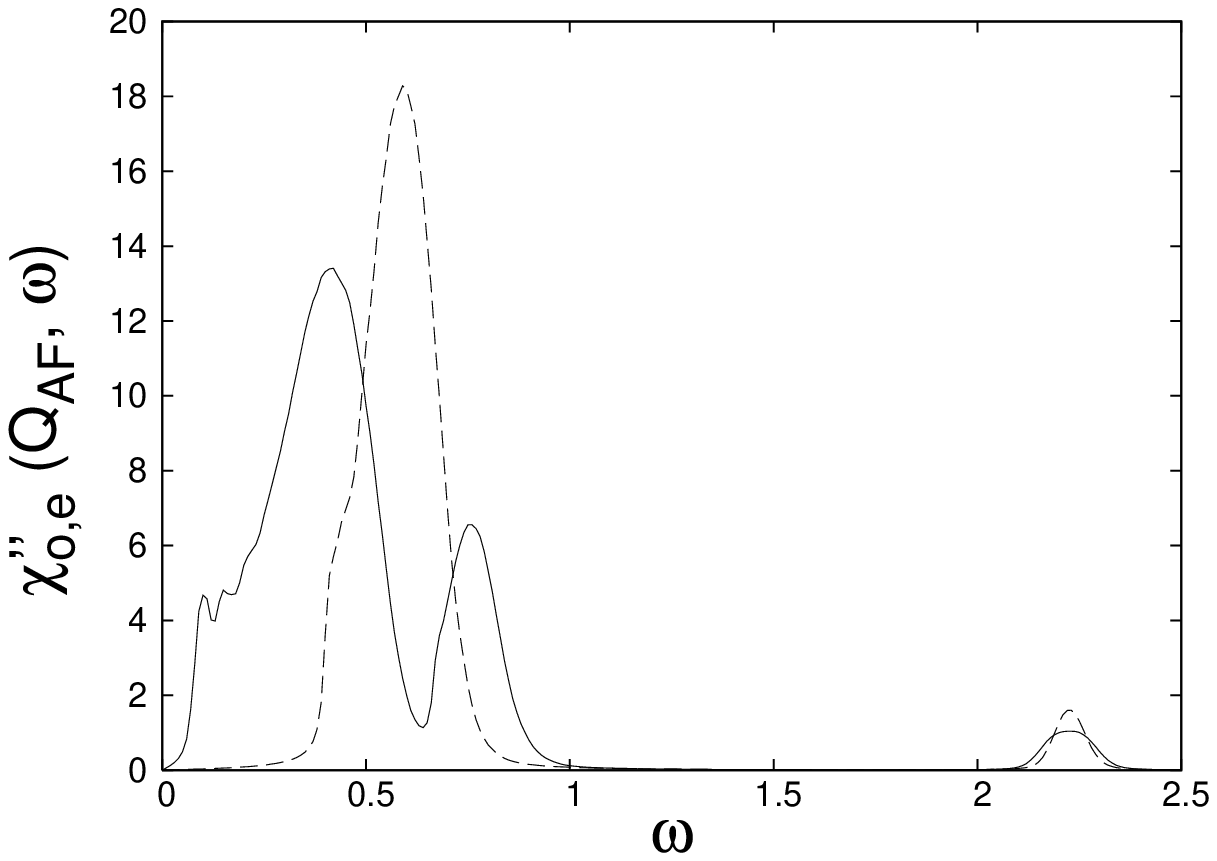}
\caption{$\protect\chi _{o,e}^{\prime \prime }(\mathbf{Q}_{\mathrm{AF}},%
\protect\omega )$ with incorporating the in-plane holon fluctuations,
simulated in the same way as in Fig.\protect\ref{fig:pipi_with_fluc}. The
odd mode: solid curve; the even mode: dashed curve.}
\label{fig:interlayer_smear}
\end{figure}

At last, we consider the effect of the holon fluctuations in the bilayer
case by using the same method introduced in Sec. \ref{sec:spin-dynamics-with}%
, and the results are plotted in Fig. \ref{fig:interlayer_smear}, in which
the solid curve is in the odd channel while the dashed curve is in the even
channel. However, we point out that the interlayer coupling should be even
more sensitive to the in-plane charge density fluctuations because the
nonlocal phase factor involved in $\Delta _{\perp }^{s}$ (see Eq.(\ref{inter}%
) in Appendix A), which is not considered in Fig. \ref{fig:interlayer_smear}%
. Generally speaking, the in-plane flux fluctuations due to the charge
fluctuations will strongly frustrate the interlayer coupling. Thus, in the
bilayer system, the former should be suppressed more, as compared to the
single-layer case, by the interlayer coupling. More studies along this line
will be conducted in future.

\section{Conclusions and Discussions}

\label{sec:discussion}

In this paper, we have studied the spin dynamics in the superconducting
state of a doped Mott insulator, which is described by the phase string
model. In this model, the spin degrees of freedom are characterized by the
bosonic RVB mean-field state, which can continuously evolve into the AFLRO
state in the zero-doping limit, where the correct spin wave excitations are
recovered.

Our study has systematically shown how the low-lying spin wave excitations
at half-filling are re-shaped into non-propagating modes in the
superconducting phase by the motion of doped holes, via the phase string
effect. We have found that the resonancelike peak near the AF wave vectors
in the superconducting phase has its dominant spectral weight, at small
doping, originated from that of the AFLRO at half-filling. That is, with the
opening up a spin gap at finite doping, the low-lying spectral weight,
including that of the condensed spinons, is pushed upward to a finite
characteristic energy of the resonancelike peak, which is linearly
proportional to the doping concentration at small doping. We have analyzed
the momentum broadening of the resonancelike peak, which decides a
characteristic spin-spin correlation length, inversely proportional to the
square root of doping concentration, or the average hole-hole distance.

Our results have also clearly illustrated that the high-energy part of the
dynamic spin susceptibility near the Brillouin zone boundary remains
essentially the same at half-filling and at small doping, with the
high-energy spin-wave signature still present in the superconducting phase.
It reflects that fact that the local and high-energy AF correlations, within
the length scale of the average hole-hole distance, have not been
drastically changed by the motion of the holes. This is in sharp contrast to
the prediction based on a Fermi-liquid-like theory, where the Fermi energy
will serve as the natural high-energy cutoff in the spin susceptibility
function. Our theory suggests that one must combine both the low-lying and
high-energy spin excitations in order to correctly understand the nature of
the spin dynamics in the high-$T_{c}$ cuprates.

The fluctuational effects beyond the mean-field theory have also been
examined. In the phase string model, the characteristic fluctuations will
come from the density fluctuations of holons, which result in the local
fluctuations of fluxes attached to holons but seen by spinons. The influence
of such fluctuational effect on the spin degrees of freedom has been found
to generally cause the broadening of the resonancelike peak in energy space,
making the emergence of some low-lying weight below the resonancelike peak.
This type of fluctuations is intrinsic and is believed to be important for
the single-layer systems like the \textrm{LSCO} compound. In particular, we
have found that the spin-spin correlation lengths of these low-lying modes
are generally longer than the `norm' one discussed in the mean-field theory.
As a consequence, the incommensurability of the spin dynamics at low
energies will show up, which is an intrinsic effect of the phase string
model but is usually not visible when the width of each peak is too broad in
momentum space, as in the `norm' mode at the resonancelike peak.

We have further investigated the interlayer effect on the spin dynamics by
considering a bilayer system. At half-filling, the spin excitation spectrum
remains the same in the odd channel as the single-layer one at low energy,
while a gap is opened up in the even channel, with the magnitude consistent
with the experiment. Then we have shown a systematic evolution of the spin
excitations, in both odd and even channels, with doping. In the
superconducting phase, the effect of the interlayer coupling is most
important for the low-lying resonancelike peak near $\mathbf{Q}_{\text{AF}}.$
A prediction for the odd channel is that there will be a second peak with a
smaller amplitude emerging at a higher energy, lying between the main
resonancelike peak in the odd channel and the peak in the even channel, near
$\mathbf{Q}_{\text{AF}}$. However, both this second peak in the odd channel
and the peak in the even channel will be sensitive to the fluctuations \emph{%
between} the two layers, which are not included in our mean-field treatment.

Finally, we point out that in the present approach, our main efforts have
been focused on the effective Hamiltonian $H_{s}$, which describes the
spinon degrees of freedom in the phase string model. The charge degrees of
freedom are described by a holon Hamiltonian, $H_{h},$ in the phase string
model, which is not considered explicitly as the holons are simply assumed
to be Bose condensed in the superconducting phase, and two degrees of
freedom are thus decoupled in this sense. But due to the mutual topological
gauge fields in the phase string model, the condensed holons can feel an
excitation from the spinon degrees of freedom and do response to it, as
discussed in Ref. \cite{muthu}.\ As a matter of fact, such a response will
result in a loose confinement of spinons to allow only the $S=$ \textrm{%
integer} types of spin excitations. We have considered the effect of $H_{h}$
within the RPA and ladder-diagram approximations and found that the results
presented in this work are not changed essentially, due to the fact that the
interactions introduced by $H_{h}$ are of logarithmic type and the spinons
excitations are localized in space in the superconducting phase. Due to the
length of the paper, we shall present these results in a separated paper.
Lastly, we remark that the superconducting phase is not stable in the phase
string model when the doping concentration is very low ($<0.04$) where the
spin ordered phase will persist, with the doped holes being localized, which
have been discussed in Ref. \cite{kou} recently.

\begin{acknowledgments}
We acknowledge helpful discussions with Z.C. Gu, T. Li, X. L. Qi, and Y.
Zhou. This work is partially supported by the grants of NSFC, the grant no.
104008 and SRFDP from MOE of China.
\end{acknowledgments}

\appendix

\section{Bosonic RVB mean-field theory for the bilayer system}

In the following, we generalize the bosonic RVB mean-field theory for the
single-layer case \cite{phasestring_general} to a bilayer system as
described by the generalized $t-J$ model (\ref{bilayer}).

We start with the phase string decomposition for the single-layer case \cite%
{phasestring_general} with explicitly introducing a layer index $l$ for each
layer $(l=1,2)$:
\begin{equation}
c_{il\sigma }=h_{il}^{\dagger }b_{il\sigma }e^{i\Theta _{il\sigma }^{\mathrm{%
string}}},
\end{equation}%
where $e^{i\Theta _{il\sigma }^{\mathrm{string}}}$ tracks the in-plane phase
string effect, defined by
\begin{equation}
\Theta _{il\sigma }^{\mathrm{string}}=\frac{1}{2}(\Phi _{il}^{b}-\sigma \Phi
_{il}^{h}),
\end{equation}%
with
\begin{equation}
\Phi _{il}^{b}\equiv \sum_{j\neq i}\theta _{i}(j)\left( \sum_{\alpha }\alpha
n_{jl\alpha }^{b}-1\right) \text{ \ \ },
\end{equation}%
and
\begin{equation}
\Phi _{il}^{h}\equiv \sum_{j\neq i}\theta _{i}(j)n_{jl}^{h}\text{ \ \ }.
\end{equation}

The exchange term in the phase string formulation reads
\begin{equation}
H_{J}^{\mathrm{bilayer}}=-\frac{J}{2}\sum_{\langle ij\rangle ,l}\left( \hat{%
\Delta}_{ij,l}^{s}\right) ^{\dagger }\hat{\Delta}_{ij,l}^{s}-\frac{J_{\bot }%
}{2}\sum_{i}\left( \hat{\Delta}_{ii,\perp }^{s}\right) ^{\dagger }\hat{\Delta%
}_{ii,\perp }^{s}\text{ }.  \label{eq:11}
\end{equation}%
where the in-plane RVB pair order parameter%
\begin{equation}
\hat{\Delta}_{ij,l}^{s}=\sum_{\sigma }e^{-i\sigma A_{ij,l}^{h}}b_{il\sigma
}b_{jl-\sigma },
\end{equation}%
and the interlayer RVB pair order parameter

\begin{equation}
\hat{\Delta}_{ii,\perp }^{s}=\sum_{\sigma }e^{-i\frac{\sigma }{2}(\Phi
_{i1}^{h}-\Phi _{i2}^{h})}b_{i1\sigma }b_{i2-\sigma }.  \label{inter}
\end{equation}%
In the single-layer case, the hopping term contributes to an additional
feedback effect \cite{phasestring_meanfield} on the spin degrees of freedom,
besides the phase string effect. But it does not qualitatively and
quantitatively change the main results of the spin dynamics. Similarly, the
interlayer hopping term is not considered here due to the same reason.

In the superconducting state, due to the holon condensation \cite%
{phasestring_meanfield}, the in-plane gauge field $A_{ij,l}^{h}$ can be
treated as describing a uniform flux [\emph{cf.} (\ref{uflux})]. On the
other hand, the phase difference between $\Phi _{i1}^{h}$ and $\Phi
_{i2}^{h} $ for two layers may be considered as a constant in the holon
condensation case, i.e. $\Phi _{i1}-\Phi _{i2}=\phi $, so that it can be
gauged away. Then it is natural to introduce the following RVB order
parameters%
\begin{eqnarray}
\Delta ^{s} &\equiv &\langle \sum_{\sigma }e^{-i\sigma
A_{ij,l}^{h}}b_{il\sigma }b_{il-\sigma }\rangle ,  \label{eq:8} \\
\Delta _{\bot }^{s} &\equiv &\langle \sum_{\sigma }b_{i1\sigma }b_{i2-\sigma
}\rangle .  \label{eq:15}
\end{eqnarray}%
The superexchange term including the interlayer coupling is thus reduced to
\begin{eqnarray}
H_{s} &=&-\frac{J\Delta ^{s}}{2}\sum_{\langle ij\rangle \sigma l}b_{il\sigma
}^{\dagger }b_{jl-\sigma }^{\dagger }e^{i\sigma A_{ij}^{h}}-\frac{J_{\bot
}\Delta _{\bot }^{s}}{2}\sum_{i\sigma }b_{i1\sigma }^{\dagger }b_{i2\sigma
}^{\dagger }+H.c.+\text{ }\mathrm{const.}  \notag \\
&&+\lambda \left( \sum_{il\sigma }b_{il\sigma }^{\dagger }b_{il\sigma
}-2(1-\delta )N\right) .
\end{eqnarray}
To diagonalize this Hamiltonian, we introduce the generalized Bogoliubov
transformation
\begin{equation}
b_{il\sigma }=\sum_{mk}\omega _{mk\sigma }(i,l)(u_{mk}\gamma _{mk\sigma
}-v_{mk}\gamma _{mk-\sigma }^{\dagger }),
\end{equation}%
where $k=\pm $. By requiring
\begin{equation}
\lbrack H_{s},\gamma _{mk\sigma }]=E_{mk}\gamma _{mk\sigma },\text{ and\ \ }%
[H_{s},\gamma _{mk\sigma }^{\dagger }]=-E_{mk}\gamma _{mk\sigma }^{\dagger },
\end{equation}%
we find
\begin{eqnarray}
(\lambda -E_{mk})u_{mk\sigma }(i,l) &=&-\frac{J\Delta _{s}}{2}%
\sum_{j=NN(i)}v_{mk-\sigma }^{\ast }(j,l)e^{i\sigma A_{ij}^{h}}-\frac{%
J_{\bot }\Delta _{\bot }^{s}}{2}v_{mk-\sigma }^{\ast }(i,l^{\prime })
\label{eq:12} \\
(\lambda +E_{mk})v_{mk\sigma }(i,l) &=&-\frac{J\Delta _{s}}{2}%
\sum_{j=NN(i)}u_{mk-\sigma }^{\ast }(j,l)e^{i\sigma A_{ij}^{h}}-\frac{%
J_{\bot }\Delta _{\bot }^{s}}{2}u_{mk-\sigma }^{\ast }(i,l^{\prime }),
\label{eq:13}
\end{eqnarray}%
where $l^{\prime }$ denotes the layer different from $l$. We obtain the
solution
\begin{equation}
u_{mk\sigma }(i,l)=u_{mk}\omega _{mk\sigma }(i,l),~v_{mk\sigma
}(i,l)=v_{mk}\omega _{mk\sigma }(i,l),
\end{equation}%
with
\begin{equation}
u_{mk}^{2}-v_{mk}^{2}=1,
\end{equation}%
and $\omega _{mk\sigma }(i,l)$ satisfies
\begin{equation}
\xi _{mk}\omega _{mk\sigma }(i,l)=-\frac{J\Delta _{s}}{2}\sum_{j=NN(i)}%
\omega _{mk-\sigma }^{\ast }(j,l)e^{i\sigma A_{ij}^{h}}-\frac{J_{\bot
}\Delta _{\bot }^{s}}{2}\omega _{mk-\sigma }^{\ast }(i,l^{\prime }).
\label{eq:14}
\end{equation}%
The spinon spectrum is given by
\begin{equation}
E_{mk}=\sqrt{\lambda ^{2}-\xi _{mk}^{2}}.
\end{equation}%
and
\begin{eqnarray}
u_{mk} &=&\sqrt{\frac{1}{2}\left( \frac{\lambda }{E_{mk}}+1\right) }, \\
v_{mk} &=&\mathrm{sgn}(\xi _{mk})\sqrt{\frac{1}{2}\left( \frac{\lambda }{%
E_{mk}}-1\right) }.
\end{eqnarray}%
According to \eqref{eq:3}, one has
\begin{eqnarray}
&&\omega _{mk\sigma }(i,l)=\frac{1}{\sqrt{2}}\left( \mathrm{sgn}(\xi
_{m})k\right) ^{l}\omega _{m\sigma }(i), \\
&&\xi _{mk}=\mathrm{sgn}(\xi _{m})\left( \left\vert \xi _{m}\right\vert +%
\frac{kJ_{\bot }\Delta _{\bot }^{s}}{2}\right) ,
\end{eqnarray}%
where $\xi _{m}$ and $\omega _{m\sigma }(i)$ are the solutions of %
\eqref{eq:3}.

Finally, the self-consistent equations of the RVB order parameters and the
Lagrangian multiplier $\lambda $ can be obtained as follows
\begin{eqnarray}
\left\vert \Delta ^{s}\right\vert ^{2} &=&\frac{1}{4NJ}\sum_{mk}\frac{\xi
_{m}\xi _{mk}}{E_{mk}}\coth \frac{\beta E_{mk}}{2}, \\
\Delta _{\bot }^{s} &=&-\frac{1}{2N}\sum_{mk}\frac{\mathrm{sgn}(\xi
_{m})k\xi _{mk}}{E_{mk}}\coth \frac{\beta E_{mk}}{2}, \\
2-\delta &=&\frac{1}{2N}\sum_{mk}\frac{\lambda }{E_{mk}}\coth \frac{\beta
E_{mk}}{2}.
\end{eqnarray}

\end{document}